\documentclass[aps,superscriptaddress,showpacs,nofootinbib,eqsecnum]{revtex4}

\def\be{\begin{equation}} \def\ee{\end{equation}} \def\bea{\begin{eqnarray}}
\def\eea{\end{eqnarray}}
\newcommand{\lsim}{\raisebox{-0.13cm}{~\shortstack{$<$ \\[-0.07cm] $\sim$}}~}

\usepackage{amsmath,lscape,epsfig} \usepackage{amsfonts} \usepackage{amsmath}
\usepackage{amssymb} \usepackage{graphicx} \usepackage{multirow}

\def\qq{\langle \bar q q \rangle} 

\begin{document}

\title{Thermodynamics and Phase Structure of the Two-Flavor
  Nambu--Jona-Lasinio Model Beyond Large-$N_c$} 

\author{Jean-Lo\"{\i}c Kneur}  \email{kneur@lpta.univ-montp2.fr}
\affiliation{Laboratoire de Physique Th\'{e}orique et Astroparticules-CNRS-UMR
  5207, Universit\'{e} Montpellier II, France} 

\author{Marcus Benghi Pinto}  \email{marcus@fsc.ufsc.br} \affiliation{Nuclear
  Science Division, Lawrence Berkeley National Laboratory, 94720 Berkeley, CA,
  USA} \affiliation{Departamento de F\'{\i}sica, Universidade Federal de Santa
  Catarina, 88040-900 Florian\'{o}polis, Santa Catarina, Brazil} 

\author{Rudnei O. Ramos} \email{rudnei@uerj.br} \affiliation{Departamento de
  F\'{\i}sica Te\'orica, Universidade do Estado do Rio de Janeiro, 20550-013
  Rio de Janeiro, RJ, Brazil} \affiliation{School of Physics and Astronomy,
  University of Edinburgh, Edinburgh, EH9 3JZ, United Kingdom}

\begin{abstract}

The optimized perturbation theory (OPT) method is applied to the $SU(2)$
version of the Nambu--Jona-Lasinio (NJL) model both at zero and at finite
temperature and/or density. At the first nontrivial order the OPT exhibits a class
of $1/N_c$ corrections which produce nonperturbative results that go beyond
the standard large-$N_c$, or mean-field approximation.  The consistency of the
OPT method with the Goldstone theorem at this  order is established, and
appropriate OPT values of the basic NJL (vacuum) parameters are obtained by
matching the pion mass and decay constant consistently.   Deviations from
  standard large-$N_c$ relations induced by OPT at this order  are derived,
for example, for the Gell--Mann-Oakes-Renner relation. Next,    the results for the
critical quantities and the phase diagram of the  model, as well as a number
of other thermodynamical quantities of interest, are obtained with OPT and
then contrasted with the corresponding results at large $N_c$.

\end{abstract}


\maketitle

\section{Introduction}

Nambu--Jona-Lasinio (NJL) models~\cite{njl} are schematic quark models useful
as a tool to understand the  physics associated with chiral symmetry and the
phase structure  in quantum chromodynamics (QCD).  NJL types of model are
extensively used in studies related to nuclear astrophysics, such as the ones
concerning neutron and quark stars, while sophisticated versions of the model
are also employed to study color superconducting phases in deconfined quark
matter in attempts to unveil the QCD phase diagram 
(for NJL model reviews, both at zero and at finite temperature and density, see
e.g. Refs.~\cite{klevansky,njlrev3,buballa}). 

Since the model does not include gluon degrees of freedom and thus cannot be
used to study confinement, its use is more suitable for the study of the 
low-temperature regime of QCD and quark matter, where the physics of confinement
is less important. However, in this regime the strong-coupling and
nonperturbative nature of the nuclear matter becomes relevant.  Many studies
using NJL-type of models have restricted themselves to the use of the
large-$N_c$ (LN) limit (where $N_c$ is the number of colors), which is also
equivalent to the Hartree  approximation~\cite{klevansky}. Although a number
of investigations of higher-order corrections beyond the Hartree approximation
have been carried out in the past (see, e.g., \cite{nloNc,oertel} in particular
for the next-to-leading $1/N_c$ calculations), these studies are in general
technically involved and not obvious,
and they may open new issues. Typically, because of the intrinsic
non-renormalisability of the model, going to higher orders requires in 
general the
introduction of new parameters, sometimes making the conclusion and
comparison with simpler LN results to  depend on extra parameters.  
It is thus of
interest to examine alternative methods able to go  beyond mean-field
approximations, to determine how they perform as compared to   the LN
approximation, and  whether they provide  both qualitatively and
quantitatively relevant corrections to the large-$N_c$ constraint.

In this paper we consider the simplest $SU(2)$ version of  the NJL model that
will be exploited beyond  the LN limit, or mean-field approximation (MFA),  by
means of the Optimized Perturbation Theory (OPT).  The OPT method (which also
goes by different names, or has many variants,
e.g. ``delta-expansion''~\cite{early}, order-dependent
mapping~\cite{zinn-justin_rev}, etc)  is notorious for allowing evaluations
beyond the MFA because of the way it modifies the ordinary perturbative expansion, giving
a nontrivial (nonperturbative) coupling dependence.  In particular, in
  models with an $O(N)$ or $SU(N)$ symmetry, an important part of
  next-to-leading $1/N$ corrections are captured at first OPT order,  although
  it is basically a different approximation scheme than the $1/N$ expansion,
  while corrections belonging formally to higher $1/N$ order are also partly
  included.  In the few studied  models where perturbative orders are
  available at high orders, the OPT turns out to improve substantially the
  convergence of ordinary perturbation, resumming the latter to some extent,
  and providing convergent sequences of approximations to some
  nonperturbative results. Examples of successful applications include the
precise determination of the critical temperature for weakly interacting dilute
Bose gases~\cite{bec}, phase diagrams for scalar theories~\cite{escalar},
Gross-Neveu (GN) types of model~\cite{prdgn2d,prdgn3d} and Yukawa
theories~\cite{yukawa},  as well as the recent precise determination of
critical dopant concentration in polyacetylene~\cite{poly}. In particular, the
precise location of the tricritical point and the mixed liquid-gas phase 
within the GN
model in 2+1 dimensions~\cite{prdgn3d} illustrates how this method can be a
powerful tool  beyond standard perturbation theory, since these important
effects were missed by the MFA and could not be precisely determined by Monte
Carlo simulations~\cite{mc}. Moreover in the $O(N)$ GN model it has been shown 
very recently~\cite{optrg} that a percentage level of
accuracy can be reached already at first OPT and $1/N$-expansion order, when 
in this case the relevant renormalization group dependence is incorporated. 
The method has also been recently applied with
success to the study of spontaneous supersymmetry
breaking~\cite{susy}. {}Finally, closer to NJL model considerations, the OPT
approach has been applied  in the past directly to the full QCD Lagrangian (at
zero temperature), in a way consistent with renormalization, obtaining, 
for instance, estimations of the quark condensate and pion decay constant 
in the chiral symmetry limit~\cite{qcdopt}.  

The OPT version adopted here is mainly indicated to nongauge theories, which
(at finite temperature) require  the method to be  extended, for example, by adding
and subtracting a hard-thermal-loop improvement that modifies the propagators
and vertices in a self-consistent way, in the so-called hard-thermal-loop
perturbation theory (HTLPT)~\cite{HTLPT}.

We will see that the OPT method, as applied at first order to the simplest 
version of the NJL model, does not introduce new effective parameters 
beyond those present in the standard MFT or LN  picture, while providing at 
the same time
  nontrivial corrections beyond LN approximation. Applying the OPT to the
NJL model actually provides a rather complete description of the thermodynamics
of the model, showing how corrections beyond LN change known results at that
level of approximation. This allows to pinpoint how important these
corrections may be in providing a reliable description for the model. By
studying different thermodynamical quantities, like the trace anomaly, specific
heat and quark susceptibility, we are also able to understand how useful these
quantities might be as indicators for precise location of the critical
points in the phase diagram, a topic of most interest today in the context of
quark-gluon phase transition and heavy-ion collision experiments.
   
This paper is organized as follows. In Sec. II we review the basic features of
the two-flavor NJL model. In Sec. III we discuss how the OPT has to be
implemented within this model. In Sec. IV we evaluate the Landau's free energy
density and its optimization is performed to the first nontrivial order in 
the OPT.  In Sec. V we establish the validity  of the Goldstone theorem at
this OPT order, and derive all necessary expressions to rederive a consistent
set of basic vacuum NJL parameters matched to the pion mass and decay
constant. In
Sec. VI we present  a series of numerical results that are relevant at 
various regimes of temperature and/or density.  The phase diagram of the 
NJL model, as well as many
relevant thermodynamical  quantities, are further explored and compared with
the corresponding LN approximation results.  Our conclusions and perspectives
are presented in Sec. VII.  Two appendixes are included to give some
relevant technical expressions and details of our calculations.

\section{The two-flavor NJL effective model for quarks}

The NJL model is described by a Lagrangian density for fermionic fields given
by~\cite{njl}

\begin{equation}
\mathcal{L}={\bar \psi}\left( i{\partial \hbox{$\!\!\!/$}}-m_{c}\right) \psi
+G\left[ ({\bar \psi}\psi)^{2}+({\bar{\psi}} i\gamma _{5}{\vec{\tau}}\psi
  )^{2}\right] ,  
\label{njl2}
\end{equation}
\noindent
where $\psi$ (a sum over flavors and color degrees of freedom is implicit)
represents a flavor isodoublet ($u,d$ types of quarks) $N_{c}$-plet quark
fields, while $\vec{\tau}$ are isospin Pauli matrices. The Lagrangian density
(\ref{njl2}) is invariant under (global) $U(2)_{\rm f}\times SU(N_{c})$ and,
when $m_{c}=0$, the theory is also invariant under chiral $U(2)_{L}\times
U(2)_{R}$. Note  that, as emphasized in
Refs. \cite{koch,buballa_stab,buballa}, the  introduction of a vector
interaction term of the form $({\bar \psi} \gamma^\nu \psi)^2$ in
Eq. (\ref{njl2}) is also allowed by the chiral symmetry and such a term can
become important at finite densities, generating a saturation mechanism
depending on the vector coupling strength that provides better matter
stability. Within the LN approximation (or MFA), the effect of such a term in
the thermodynamical potential is to produce a shift on the chemical
potential. However, this term will not be considered here.

Due to the quadratic fermionic interaction, the theory is nonrenormalizable
in 3+1 dimensions ($G$ has dimensions of $\mathrm{eV}^{-2}$), meaning that
divergences appearing at successive perturbative orders cannot be all
eliminated  by a consistent redefinition of the original model parameters
(fields, masses, and couplings). The renormalizability issue arises during the
evaluation of momentum integrals associated with loop Feynman graphs in a
perturbative expansion and, in the process, one usually employs regularization
prescriptions (e.g. dimensional regularization, sharp cutoff, etc) to
formally isolate divergences. However, the procedure introduces
\textit{arbitrary} parameters with dimensions of energy that do not appear in
the \textit{original} Lagrangian density.  
Within the NJL model a sharp cut off ($\Lambda$) is preferred and since the
model is nonrenomalizable, one has to  fix $\Lambda$ to a value related to the
physical spectrum under investigation. This strategy turns the 3+1 NJL model
into an effective model, where $\Lambda$ is treated as a parameter, as usual
in effective nonrenormalizable field theory models.  The experimental values
of quantities such as the pion mass ($m_{\pi}$) and the pion decay constant
$(f_{\pi})$ are used to fix both, $G$ and $\Lambda$. An interesting
alternative regarding  regularization within the  NJL model is presented in
Ref.~\cite{gastaoNJL} (where explicit evaluation of divergent integrals is avoided 
by assuming in intermediate steps only general symmetry properties of the regularization, 
such that the finite parts are integrated in a way independent of the regularization).

A second important issue regards the fact that, when $m_c = 0$, the quark
propagator brings unwanted infrared divergences, meaning that the evaluations
have to be carried out in a \textit{nonperturbative} fashion.  Moreover, very
often physical quantities (like the self-energy) appear as powers of the
dimensionless quantity $G \Lambda^2$, which is greater than unity, preventing
any possibility of calculations via standard perturbative methods.

In analytic nonperturbative evaluations, one can consider one-loop
contributions dressed by a fermionic propagator, whose effective mass, $M$,
is determined in a self-consistent way. This approximation is known under
different names, for example, the Hartree, LN or mean-field approximation.  
To obtain the effective potential (or Landau free energy density), 
$\mathcal{F}$, for
the quarks, it is convenient to consider the bosonized version of the NJL,
which is easily obtained by introducing auxiliary fields ($\sigma ,{
  \vec{\pi}}$) through a Hubbard-Stratonovich type of transformation. Here,
$\mathcal{F }$ is evaluated using the LN approximation, which is equivalent to
the MFA. Then, to introduce the auxiliary bosonic fields and to render the
theory more suitable for use of the LN approximation, it is convenient to use
$G\rightarrow \lambda /(2N_{c})$ and to formally treat $N_{c}$ as a large
number, which is set to the relevant value, $N_{c}=3$, at the end of the
evaluations. One then has

\begin{equation}
\mathcal{L}={\bar \psi}\left( i{\partial \hbox{$\!\!\!/$}}-m_{c}\right) \psi
-{\bar \psi}(\sigma +i\gamma _{5}{\vec{\tau}\cdot }{\vec{\pi}} )\psi
-\frac{N_{c}}{2\lambda }(\sigma ^{2}+{\vec{\pi}}^{2}).
\label{njlboson}
\end{equation}
\noindent
At finite temperature and density the model can be studied in terms of the
grand partition function, defined as usual by

\begin{equation}
Z(\beta ,\mu )=\mathrm{Tr}\exp \left[ -\beta \left( H-\mu Q\right) \right]
\;,  \label{Zbetamu}
\end{equation}
\noindent
where $\beta$ is the inverse of the temperature, $\mu$ is the chemical
potential for both flavors, $H$ is the Hamiltonian corresponding to
Eq. (\ref{njlboson}) and $Q= \int d^3 x \bar{\psi} \gamma_0 \psi$ is 
the mean baryon charge.

\section{Interpolation of the NJL Model}

To implement the OPT within the NJL model one follows the prescription used in
Refs.~\cite{prdgn2d,prdgn3d} by first interpolating the original four-fermion
version in terms of a fictitious parameter $\delta$, which is the new
expansion parameter. For a long, but far from complete list of references on
this and related methods, see \cite{early,linear}. See also
\cite{zinn-justin_rev} for a recent review.  According to this prescription
the deformed Lagrangian density for the NJL model in terms of the auxiliary
fields becomes 

\begin{equation}
\mathcal{L}=\bar{\psi}\left[ i{\partial \hbox{$\!\!\!/$}}-m_{c}-\delta \left(
  \sigma +i{\gamma }_{5}{\vec{\tau}\cdot \vec{\pi}} \right) -\eta \left(
  1-\delta \right) \right] {\psi }-\delta \frac{N_{c} }{2\lambda }\left(
\sigma ^{2}+{\vec{\pi}}^{2}\right) .  
\label{delta12}
\end{equation}
\noindent
In order to discuss issues related to chiral symmetry breaking (CSB) and
Goldstone's theorem, it is useful to temporarily consider the chiral limit of
the original theory by setting $m_{c}=0$. Note then that the actual chiral
limit and original Lagrangian is recovered for $\delta\to 1$, and that $\eta$
at this stage is an arbitrary mass parameter, as usual in the OPT method.
Now, a well-established result concerning the OPT evaluation of the free
energy density (or effective potential), $\mathcal{F}$, in the LN limit, shows
that $\bar{\eta}$  becomes exactly the classical value of the background
fields, so that both approximations coincide in this limit. {}For example, if
one sets $\pi_{i}=0$ in Eq.~(\ref{delta12}), the NJL becomes analogous to the
GN  model, displaying discrete CSB. In this case, numerous applications show
that, when considering the LN limit, we obtain $\bar{\eta}=\sigma_c$. However,
within the NJL model, apart  from the scalar channel, one also has to deal with the
pseudoscalar channel. This situation was addressed in detail in
Ref.~\cite{firstOPT}, where it was shown that the interpolation mass
parameter, $\eta$, can be extended to account for arbitrary mass   parameters
in the pseudoscalar direction. This can be accomplished by redefining $\eta$
in Eq. (\ref{delta12}) such that

\begin{equation}
\eta \to \eta +i{\gamma }_{5}{\vec{\tau}\cdot \vec{\beta}},   
\label{eta12}
\end{equation}
\noindent
implying in the most general case, four mass parameters, $\eta$ and the three
components of $\vec{\beta}$, to be fixed by a well-defined prescription
(optimization) to determine them.  However, as the Landau free energy (or
equivalently, the effective potential) is concerned, only the fluctuations  in
the scalar direction become relevant when only the scalar field $\sigma$
acquires a nonzero  vacuum expectation value ($\langle \sigma \rangle \equiv
\sigma_c$ by slight abuse of notation).    In other words, one assumes from
now on that $ \langle \pi_i \rangle =0$, which can be shown to
imply~\cite{firstOPT} within the OPT that $\bar{\beta}_{i}=0$\footnote{A more
  general scenario with extra variational parameters $\beta_i \ne0$ could be
  relevant to address the further breaking of the remaining $SU(2)_{L+R}$
  symmetry, a case that is not considered here.}.   Taking this simplest
solution, one needs only to consider the simplest variational interpolation
involving only one  mass parameter $\eta$, as explained below. 

Once the free energy density  $\mathcal{F}$ is evaluated to a given  order $k$
in the OPT, the optimization procedure used to fix the arbitrary mass
dependence follows by a specific prescription~\cite{prdgn3d}, such as the
principle of minimal sensitivity (PMS)~\cite{pms},

\begin{equation}
\left. \frac{d\mathcal{F}^{(k)}}{d\eta }\right\vert _{\bar{\eta},\delta
  =1}=0\;.
\label{pms12}
\end{equation}

\section{Optimized Free Energy Density}

\begin{figure}[tbh]
\vspace{0.5cm} \epsfig{figure=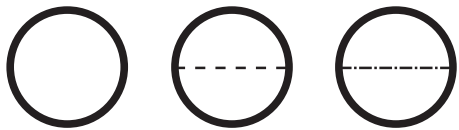,angle=0,width=6cm}
\caption{Diagrams contributing to $\mathcal{F}\left( {\hat{\protect\eta}}
  \right) $ to order $\protect\delta$. The thick continuous fermionic lines
  represent $\ {\hat{\protect\eta}}$-dependent terms which must be further
  expanded. The dashed line represents the $\protect\sigma $ and
  the $\protect\pi $ is represented by the dashed-doted line. Note that both are 
nonpropagating  
  at this level of approximation. The first
  diagram contributes with $1/N_{c}^{0}$, the second and third diagrams (of order
  $\protect\delta $) contribute with $1/N_{c}$.}
\label{effpot}
\end{figure}

To order $\delta$, Landau's free energy density is given by the diagrams shown
in {}Fig~\ref{effpot}. They are evaluated using OPT dressed propagators, where
the mass term, using a compact notation, is given by

\begin{equation}
\hat{\eta}=\eta +m_c  -\delta \left[ \eta -\left( \sigma +i{ \gamma
  }_{5}{\vec{\tau}}\cdot {\vec{\pi}}\right) \right]\;,
\label{hateta}
\end{equation}
\noindent
whose form is useful to produce results both for $m_c\ne 0$ and for   the
chiral symmetric limit $m_c= 0$. 

In the {}Feynman diagrams displayed in {}Fig. \ref{effpot}, the free energy density
in the $\sigma_c$ direction reads 

\begin{eqnarray}
\frac{\mathcal{F}}{N_{c}} &=&\frac{\sigma_c^{2}}{2\lambda
}+i \int \frac{d^{4}p}{\left( 2\pi \right) ^{4}}\mathrm{Tr}\,
{\ln
  \left( p{\hbox{$\!\!\!/$}\mathbf{-}}\eta - m_c \right) }+i
\int \frac{d^{4}p}{\left( 2\pi \right) ^{4}}\mathrm{Tr}\left( \frac{\eta
  -\sigma_c }{p{\hbox{$\!\!\!/$}
    \mathbf{-}}\eta -m_c}\right)  \notag \\ &&+\frac{1}{2}\frac{\delta \lambda
  }{N_c}\int \frac{d^{4}p}{\left( 2\pi \right) ^{4} }\int
\frac{d^{4}q}{\left( 2\pi \right) ^{4}}\mathrm{Tr}\left( \frac{1}{p{
    \hbox{$\!\!\!/$}}-\eta -m_c}\right) \left( \gamma _{5}\tau_i \frac{1}{q{
    \hbox{$\!\!\!/$}}-\eta -m_c}\gamma _{5}\tau_i \right) \notag
\\ &&-\frac{1}{2}\frac{\delta \lambda }{N_{c}}\int
\frac{d^{4}p}{\left( 2\pi \right) ^{4}}\int \frac{d^{4}q}{\left( 2\pi \right)
  ^{4}}\mathrm{Tr} \left( \frac{1}{p{\hbox{$\!\!\!/$}}-\eta -m_c}\right)
\left( \frac{1}{q{ \hbox{$\!\!\!/$}}-\eta -m_c }\right) \;. 
\label{delta4}
\end{eqnarray}
\noindent
The traces in Eq. (\ref{delta4}) are over flavor and Dirac matrix indices.  
Then, after some algebra,  one arrives at

\begin{eqnarray}
\frac{\mathcal{F}}{N_{c}} &=&\frac{\sigma_c^{2}}{2\lambda }
+ 2i N_{\rm f}\int \frac{d^{4}p}{\left( 2\pi \right) ^{4}}{\ln }\left[ -p^{2}+
(\eta+m_c)^2 \right] \nonumber \\ &-&  4 i \delta  N_{\rm f} \int \frac{
  d^{4}p}{\left( 2\pi \right) ^{4}}\frac{(\eta+m_c) \left( \eta -\sigma_c
  \right) }{-p^{2}+(\eta+m_c)^2 }  \notag \\  &&-2(n_\pi+1)\, \frac{\delta
  \lambda  N_{\rm f}}{N_{c}} \left[ \int \frac{d^{4}p}{\left( 2\pi \right)
    ^{4}}\frac{p_{0}}{-p^{2}+(\eta+m_c)^2  }\right] ^{2}   \nonumber \\ &+& 2
(n_\pi-1) \,\frac{\delta \lambda N_{\rm f}}{N_{c}} \:(\eta+m_c)^2  \:\left[
  \int \frac{d^{4}p}{\left( 2\pi \right) ^{4}}\frac{1}{-p^{2}+(\eta+m_c)^2
  }\right]^{2}  \;,
\label{delta5}
\end{eqnarray}
\noindent
where $n_\pi$ represents the number of pseudoscalars. It is interesting to see
the type of loop contributions contained in Eq. (\ref {delta5}). {}First let
us consider the $N_c^0$ contributions. The second term in Eq. (\ref {delta5})
corresponds to a gas of free fermions whose mass has been dressed, while
the third term represents tadpole-type of contributions, proportional to the
quark condensate, $\langle {\bar \psi} \psi \rangle$.  The $1/N_c$
contributions, given by the two last lines in Eq. (\ref{delta5}), are proportional to  
$(2\langle { \psi}^+ \psi \rangle^2-\langle {\bar \psi} \psi \rangle^2)$ where $\langle
{\psi}^+ \psi \rangle=\langle {\bar \psi} \gamma^0 \psi \rangle$ represents
the total quark number density. Note also that for the $U(1)$ version of the
model, $N_{\rm f}=1$ and $n_\pi=1$, so that the last term does not contribute
and the OPT will bring $1/N_c$ corrections  only at finite density.   However,
for the case under study here $n_\pi=3$ and finite $N_c$  corrections are
expected to occur at any temperature and density regime. 

All our momentum integrals are to be interpreted in the  Matsubara
finite-temperature formalism, 

\begin{equation}
\int \frac{d^{4}p}{\left( 2\pi \right)^{4}} \equiv \frac{i}{\beta}
\sum_{n=-\infty}^{+\infty} \int \frac{d^{3}p}{\left( 2\pi \right) ^{3}}\;,
\end{equation}
and quadrimomenta given as $p=(i\omega_n+\mu,{\bf p})$, with
$\omega_n=(2n+1)\pi T,\;n=0,\pm 1,\pm 2, \ldots$, are the Matsubara
frequencies for fermions. The relevant Matsubara's sums are given in 
Appendix A. One can now fix $\delta =1$ and optimize $\mathcal{F}$ using the
PMS relation, given by Eq. (\ref{pms12}).  Application of the PMS condition to
Eq. (\ref{delta5}) gives

\begin{eqnarray}
 \frac{d\mathcal{F}}{d\eta }\Bigr|_{\bar{\eta},\delta =1}  &=& {-}4i
 N_{c}N_{\rm f}\int \frac{d^{4}p}{\left( 2\pi \right) ^{4} }\frac{\left(
   \bar{\eta}-\sigma_c \right) }{-p^{2}+({\bar{\eta}+m_c)^{2}} }  \notag
 \\ &&{-}4i N_{c}N_{\rm f}\left[ (\bar{\eta}+m_c)\left( \bar{\eta}-\sigma_c
   \right) \right] \frac{d}{d\eta }\left[ \int \frac{d^{4}p}{\left( 2\pi
     \right) ^{4}}\frac{1}{-p^{2}+(\bar\eta+m_c)^2 }\right]\Bigr|_{\bar{\eta}}
 \notag \\ &&-16\lambda N_{\rm f}\left[ \int \frac{d^{4}p}{\left( 2\pi \right)
     ^{4}}\frac{ p_{0}}{-p^{2}+({\bar{\eta}+m_c)^{2}} } \right] \frac{d}{d\eta
 }\left[ \int \frac{d^{4}p}{\left( 2\pi \right)^{4}}\frac{
     p_{0}}{-p^{2}+(\bar\eta+m_c)^2 }\right]\Bigr|_{\bar{ \eta}}  \notag
 \\ &&+8\lambda N_{\rm f} (\bar{\eta}+m_c)^2 \; \int \frac{d^{4}p}{\left( 2\pi
   \right)^{4}} \frac{1}{-p^{2}+(\bar{\eta}+m_c)^{2}} \frac{d}{d\eta }\left[
   \int \frac{d^{4}p}{\left( 2\pi \right)
     ^{4}}\frac{1}{-p^{2}+(\bar\eta+m_c)^2 }\right]\Bigr|_{\bar{ \eta}} \notag
 \\  &&+8\lambda N_{\rm f} (\bar{\eta}+m_c) \left[ \int \frac{d^{4}p}{\left(
     2\pi \right) ^{4}} \frac{1}{-p^{2}+(\bar{\eta}+m_c)^{2}} \right]^2 =0\;.
\label{delta6}
\end{eqnarray}

Now, considering Eq.~(\ref{delta5}) at finite temperature and chemical
potential, we can write it in the more compact form

\begin{eqnarray}
{\mathcal{F}}& =&\frac{\sigma_c^{2}}{4 G}-2N_{\rm f}N_c I_1 (\mu,T) +2 \delta
N_{\rm f}N_c(\eta+m_c) \left( \eta-\sigma_c \right) I_2(\mu,T) \notag\\  && +
4\delta G N_{\rm f}N_{c} \:I_3^{2}(\mu,T) - 2\delta G  N_{\rm
  f}N_{c}\,(\eta+m_c)^2 I^2_2(\mu,T)  \;,
\label{landauenergy}
\end{eqnarray}
\noindent
where we have replaced $\lambda \to 2 G N_c$.  In this equation we have
defined, for convenience, the following basic relevant integrals:

\begin{equation}
I_1(\mu,T) = \int \frac{d^{3}p}{\left( 2\pi \right) ^{3}}\left\{ E_p + T\ln
\left[ 1+e^{-\left( E_{p}+\mu \right) /T}\right] +T\ln \left[ 1+e^{-\left(
    E_{p}-\mu \right) /T}\right]  \right\} \;,
\label{defI1}
\end{equation}

\begin{equation}
I_2(\mu,T) = \int \frac{d^{3}p}{\left( 2\pi \right)^{3}} \frac{1}{E_{p}}\left[
  1-\frac{1}{e^{\left( E_{p}+\mu \right) /T}+1}-\frac{1}{e^{\left( E_{p}-\mu
      \right) /T}+1}\right]\;,
\label{defI2}
\end{equation}
and

\begin{equation}
 I_3(\mu,T) =\int \frac{d^{3}p}{\left( 2\pi \right)^{3}} \left[
   \frac{1}{e^{\left( E_{p}-\mu \right) /T}+1}-\frac{1}{e^{\left( E_{p}+\mu
       \right) /T}+1}\right]\;,
\label{basicint}
\end{equation}
\noindent
where $E_{p}^{2}={\bf p}^2+(\eta+m_c)^{2}$.  The divergent integrals occurring
at $T=0$ and $\mu=0$ are

\begin{eqnarray}
I_1(0,0) &=& \int \frac{d^{3}p}{\left( 2\pi \right) ^{3}} E_{p} \nonumber
\\ &=& \frac{1}{32\pi ^{2}} \left\{ (\eta+m_c)^{4} \ln \left[ \frac{ \left(
    \Lambda + \sqrt{ \Lambda^{2}+(\eta+m_c)^{2}} \right)^{2}}
  {(\eta+m_c)^{2}}\right] -
2\sqrt{\Lambda^{2}+(\eta+m_c)^{2}}\left[2\Lambda^{3}+ \Lambda (\eta
  +m_c)^{2}\right] \frac{}{}\right \}
\label{I1div}
\end{eqnarray}
and

\begin{equation}
I_2(0,0) = \int \frac{ d^{3}p}{\left( 2\pi \right)^{3}}
\frac{1}{E_{p}}=\frac{1}{4\pi^2} \left\{ \Lambda
\sqrt{\Lambda^{2}+(\eta+m_c)^{2}} -\frac{(\eta+m_c)^{2}}{2} \ln \left[
  \frac{\left[ \Lambda+\sqrt{\Lambda^{2}+
        (\eta+m_c)^{2}}\right]^{2}}{(\eta+m_c)^{2}}\right] \right\}\;,
\label{I2div}
\end{equation}
\noindent
where in both Eqs. (\ref{I1div}) and (\ref{I2div}) we have introduced a  sharp
noncovariant three-dimensional (3D) momentum cutoff $\Lambda$, as is most commonly done in NJL
calculations in a medium. 
In fact the two-loop graphs from Fig \ref{effpot},
  relevant  for the evaluation of the free energy in Eq.(\ref{delta4}) (or
  similar graphs, as we will see, relevant to evaluate the pion mass and decay
  constant, $m_\pi$ and $f_\pi$), involve only contributions of ``auxiliary''
  pion and sigma fields.  As a result the former reduce to simple one-loop
  contributions squared, which are automatically finite when a single  
  $\Lambda$ cutoff parameter is used~\footnote{In contrast in the genuine $1/N_c$ corrections,  (dressed)
    meson propagators in graphs similar to {}Fig.~\ref{fig2loop} involve
    integration over two independent momenta, and an independent cutoff
    $\Lambda_M$ parameter is sometimes introduced to regularize the meson
    loops~\cite{oertel}, so that more data are 
   needed to fix all the model parameters}. The OPT method provides a  nontrivial relation between the
  variational mass $\eta$  and the coupling $\lambda$, but at first order this
  amounts to peculiar mass insertions into essentially one-loop calculations
  (eventually resummed in the so-called random phase approximation (RPA), when
  we consider the pion mass and its connection with the Goldstone theorem
  realization). Thus, the cutoff $\Lambda$ in integrals like
  Eq. (\ref{I2div}) plays a similar role as in the Hartree (or LN) approximation,
  though its value will be modified by OPT corrections when a consistent
  matching of OPT expressions to the pion data  is done, as we will examine in
  the next section.\\ 
  Here, we impose the cutoff only for the vacuum term, since the finite temperature has
a natural cutoff in itself specified by the temperature. This choice of
regularization, which allows for the Stefan-Boltzmann limit  to be reproduced
at high temperatures, is sometimes preferred in the
literature~\cite{fukushima}.

In order to perform our evaluations we need to consider the general PMS
equation (\ref{delta6}), which can be conveniently expressed  in the form

\begin{equation}
 \left\{ \left[\eta - \sigma_c - 2 (\eta+m_c) G \, I_2\right] \left[ 1 +
   (\eta+m_c) \frac {d}{d \eta} \right] I_2 + 4 G \, I_3 \frac{d}{d \eta} I_3
 \right\}_{\eta = {\bar \eta}}=0 \,.
\label{simplePMS}
\end{equation}
\noindent
Since we are mainly interested in the thermodynamics, one basic quantity of
interest is the thermodynamical potential, ${\cal V}$, whose relation to 
the free energy is given by ${\cal V}= {\cal F}({\bar \sigma}_c)$. The order
parameter, ${\bar \sigma}_c$, is determined from the gap equation generated by
minimizing ${\cal F}$ with respect to the classical field, ${\sigma}_c$. 
{}From Eq. (\ref{landauenergy}) we obtain that

\begin{equation}
 {\bar \sigma}_c = 4 G N_{\rm f} N_c (\eta+m_c) I_2\;.
\label{simpleGAP}
\end{equation}
\noindent
A rather nice analytical result emerges in the case where $\mu=0$, since in
this case the last term of Eq. (\ref{simplePMS}) vanishes and one obtains that\footnote{Note that
the factor $2$ in Eq. (\ref{etasig1}) is actually $n_\pi-1$ with $n_\pi=3$, as can be traced from the
last line of Eq. (\ref{delta5}). This illustrates, as already mentioned, that those first OPT order
$1/N_c$ corrections vanish in the $U(1)$ case for $\mu=0$.}

\begin{equation} 
 {\bar \eta} =  \sigma_c + 2 G ({\bar \eta}+m_c) I_2\;,
\label{etasig1}
\end{equation}
\noindent
from which  follows the simple relation (for $\mu=0$)
\begin{equation}
{\bar \eta} =  {\bar \sigma_c} {\cal G}(N) \;.
\label{GAPPMS}
\end{equation}
The $1/N_c$-dependent term
\begin{equation}
 {\cal G}(N)= \left ( 1+ \frac {1}{2 N_{\rm f} N_c} \right ) \,\,,
\end{equation}
then corrects the LN relation $\bar\eta=\bar\sigma_c$~\cite{firstOPT}.

\section{OPT Mass gap, Goldstone theorem, and basic vacuum NJL parameters}

In this section we shall derive some important steps for subsequent
calculations at finite temperature and density. We first establish the OPT
corrections to some basic expressions relevant to determining the NJL
parameters, as compared with the corresponding large-$N_c$ results.   The NJL
Lagrangian density represents an effective model whose parameters should be
determined from data (most conveniently for vacuum quantities at $T=\mu=0$)
before one attempts to make predictions for other physical quantities. In order
to make  a sensible comparison quantifying the size of the corrections beyond
LN approximation induced by our approach, 
we shall first derive consistently the basic
parameters from data for both the LN and the OPT cases.

\subsection{The vacuum mass gap }

A first crucial step is to examine for the case $T=0$ and $\mu=0$ the
mechanism through which the quark masses shift from their current value,
$m_c$, to the effective value, which at the present level of approximation, is
given by  $M\equiv m_c + {\bar \sigma}_c$. Using Eqs.  (\ref{simplePMS}),
(\ref{simpleGAP}) and (\ref{GAPPMS}), one can write the  OPT self-consistent
gap equation:

\begin{equation} 
M^{\rm OPT}_q=m_c+{\bar \sigma}_c^{\rm OPT}= m_c + 4 G N_{\rm f}N_c  \; {\cal
  M} I_2(\mu=0,T=0)\Bigr|_{\bar{\eta}+m_c = {\cal M}} \,,
\label{optMassVac}
\end{equation}
where we have defined for convenience 

\begin{equation}
{\cal M} \equiv \bar\eta +m_c =  M^{\rm OPT}+ \frac{M^{\rm OPT}-m_c}{2 N_{\rm
    f} N_c}\;,
\label{calM}
\end{equation}
\noindent
while the LN result is

\begin{equation} 
M^{\rm LN}_q=m_c+{\bar \sigma}_c^{\rm LN}= m_c + 4 G N_{\rm f}N_c M^{\rm
  LN}_q I_2(\mu=0,T=0)\Bigr|_{\bar{\eta}+m_c \to M^{\rm LN}_q}\,.
\end{equation}

\subsection{Goldstone theorem and the OPT pion mass}

In order to derive the OPT corrections to the pion mass and decay constant, an
important and related feature is to examine whether and how the Goldstone
theorem manifests within our framework. In the Hartree (LN) approximation, it
is well known that the NJL model exhibits the massless pion poles in the
chiral limit~\cite{klevansky,buballa}, when the quark-antiquark
$T$-matrices are considered in the geometrically resummed approximation.  Similarly, here we
will see how the OPT at first order generalizes this  result. 

The Goldstone properties of the pion are exhibited more simply by taking
$q^2=0$ ($q$ denoting the external momentum of the quark-antiquark scattering
matrix),  in this case it just amounts to showing that the resummed pion
propagator has a pole, while the complete inverse pion propagator for $q^2\ne
0$ will behave as $\sim 0 +{\cal O}(q^2)$.  In what follows we consider some
expressions in Minkowski space with  covariant four-momentum for simplicity,
working under cover of a covariant regularization like Pauli-Villars one
typically, as is usual in  most NJL standard treatments of the Hartree
approximation.  It is first useful to recall how the Goldstone theorem is
realized in the LN approximation. {}For that purpose we define, in Minkowski
space, the basic integral appearing in the  gap equation~\footnote{As compared
  to our conventions in Eq. (\ref{I2div}), note that $2i\,I_G(m) =
  I_2(0,0)$.}:  

\begin{equation} 
I_G(m) = \int \frac{d^4p}{(2\pi)^4} \frac{1}{p^2-m^2} \;,
\end{equation}
\noindent
where $m$ is the relevant quark mass to be specified, depending on the
approximation level (i.e. $m\to M^{\rm LN}_q$ at large-$N_c$).  Next, the
one-loop pion self-energy has the well-known expression (see
e.g. \cite{buballa}):  

\begin{equation} 
i\Pi^{(1)}(q^2) \delta^{ij} =  N_c\int \frac{d^4 p}{(2\pi)^4} \:{\rm Tr}
\,\left [ \frac{i}{p{
      \hbox{$\!\!\!/$}}-m}(i\tau_i \gamma_5) \frac{i}{p{ \hbox{$\!\!\!/$}} +q{
      \hbox{$\!\!\!/$}}-m} (i\tau_j \gamma_5)\right] \;,
\label{Pipion}
\end{equation} 
\noindent
where the trace is over flavor and Dirac matrix indices only.
After some algebraic manipulations, Eq. (\ref{Pipion}) may be cast into the form 
\begin{equation}
\Pi^{(1)}(q^2) = 2i N_{\rm f} N_c \left[2I_G(M^{\rm LN}_q) -q^2
  I(q^2)\right]\;,
\label{pi1}
\end{equation}
\noindent
with 

\begin{equation} 
I(q^2) = \int  \frac{d^4 p}{(2\pi)^4}\;\frac{1}{(p^2-m^2)[(p+q)^2-m^2]} \;,
\end{equation}
\noindent
where we took the chiral limit  everywhere for the moment, and $m$ is to
be replaced by the appropriate value of the mass gap, $m\to M^{\rm LN}_q$.
The geometrically resummed (inverse) pion propagator is then given by 

\begin{equation} 
1-2G \Pi^{(1)}(0)\;.
\label{pionpro}
\end{equation} 
\noindent
On the other hand the (large-$N_c$) gap equation reads  

\begin{equation} 
M^{\rm LN}_q= 8 i G N_{\rm f} N_c M^{\rm LN}_q \: I_G(M^{\rm LN}_q)\;,  
\end{equation} 
\noindent
or, equivalently, 

\begin{equation} 
1-8 i G N_{\rm f} N_c \: I_G(M^{\rm LN}_q) =0\;,
\label{gap1}
\end{equation}
\noindent
implying that  (\ref{pionpro}) is also zero at $q^2=0$ upon use of the
gap equation. 

\begin{figure}[h!]
\vspace{0.5cm}    \epsfig{figure=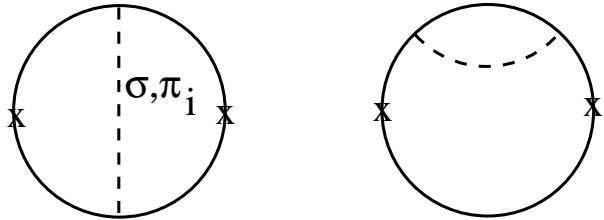,angle=0,width=8cm}
\caption{Two-loop contributions to the two-point functions relevant to the
  calculation of the pion mass and decay constant $m_\pi$ and $f_\pi$ calculations.  The crosses
  represent appropriate vertices  (e.g. $i\gamma_5 \tau_i$ or $i\gamma_\mu
  \gamma_5 \tau_i$, see the main text). Note that both the $\sigma$ and $\pi$ are non-propagating  
  at this level of approximation.}
\label{fig2loop}
\end{figure}

At first OPT order, we have derived the improved effective potential as
given by Eq. (\ref{delta5}), involving two-loop contributions with $\sigma$
and $\pi_i$ exchange according to {}Fig. \ref{effpot}.  Accordingly, for
consistency, the pion inverse propagator should be calculated at the same
order in $\delta \lambda$, and the $\delta$ dependence as induced by
Eq.~(\ref{hateta}) should be carefully expanded within the one-loop
contributions (of course taking $\delta =1$ again at the end of the
calculations). The relevant two-loop contributions are shown in
{}Fig. \ref{fig2loop}, where it should be noted that the second type of (mass
insertion) diagram is consistently generated from the $\delta$-expansion of
the one-loop diagram, which amounts to put in all one-loop expressions $m\to
\bar \eta+m_c$, where now $\bar\eta$ is defined from Eqs. (\ref{GAPPMS}).
\noindent
\noindent
Therefore, only the first type of (vertex correction) diagram needs to be
calculated, which is done in detail in Appendix B,  giving the result
Eq. (\ref{pi2sum}) for arbitrary external momentum $q^2$.  {}For $q^2\to 0$
Eq.~(\ref{pi2sum}) simplifies considerably to: 

\begin{equation} 
\Pi^{(2),ps}(0) =  -8 G N_{\rm f} N_c n_\pi\,I^2_G(m)\;, 
\end{equation} 
\noindent
having set as usual $\lambda=2N_c G$ (and $n_\pi=3$).

Next, the same type of diagram as evaluated above, but now with a  $\sigma$
scalar exchange gives a similar result, for $q^2=0$, except for an overall
minus sign (and $n_\pi \to 1$). In summary the perturbative expansion of the
inverse pion propagator at two-loop order (and $q^2\to 0$) reads, 

\begin{equation} 
1-2G\,[\Pi^{(1)}(0) +\Pi^{(2)}(0)]\;,
\end{equation} 
\noindent
or, more explicitly, 

\begin{equation} 
1- 8 i G N_{\rm f} N_c \: I_G(m) +16 N_{\rm f} N_c (n_\pi-1)\:G^2\:
I^2_G(m)\;.
\label{gold2}
\end{equation} 
\noindent
At first OPT order, the gap equation is modified in a nonperturbative way,
giving the  relation (\ref{optMassVac}). However, this result has to be
perturbatively expanded to order $\lambda$ (or $G$ equivalently) to see the
cancellations occurring at perturbative level with the extra two-loop vertex
contribution contained in Eq. (\ref{gold2}).  We obtain the required perturbative
expansion simply by taking Eq. (\ref{simpleGAP}) for $m_c=0$, 

\begin{equation}
M_{q}^{\rm OPT}\equiv \bar\sigma_c = 8 i G N_{\rm f} N_c \bar\eta \,I_G
(\bar\eta)\;, 
\end{equation}
\noindent
together with the PMS equation (\ref{etasig1}), 

\begin{equation} 
\bar\eta  = \bar\sigma_c +4 i\, G \bar\eta I_G(\bar \eta)\;,  
\end{equation} 
\noindent
and iterating once to get

\begin{equation} 
M_{q}^{\rm OPT}\equiv \bar\sigma  = 8 i G N_{\rm f} N_c \bar\sigma_c \,I_G
(\bar\eta)\: \left[1+4i\,G\,I_G(\bar \eta)+{\cal O}(G^2)\right] \;,
\end{equation}   
\noindent
such that the 2-loop (perturbatively expanded) OPT gap equation reads

\begin{equation} 
1- 8 i G N_{\rm f} N_c \: I_G(\bar\eta) +32 N_{\rm f} N_c \:G^2\: I^2_G(\bar\eta) =
0\;.
\label{gap2}
\end{equation} 
\noindent
This shows that Eq.(\ref{gold2}) also gives zero, upon identifying in the
latter $m \to \bar\eta$, since all masses within the integrands are $\bar\eta$
at the OPT level (in the chiral limit).

Recovering $m_c\ne0$, and defining the pion mass as the pole of the propagator
for $q^2\equiv m^2_\pi$, whose expression is given by Eq. (\ref{pi2sum}),
we obtain a final expression at the first OPT order for the relation between the pion mass,
$m_c$, involving also the other NJL parameters $G$, $\Lambda$, and $m_c$: 

\begin{equation} 
\frac{m_c}{M^{\rm OPT}_q} = 4 G N_f N_c m^2_\pi \left\{ -i\: I(m^2_\pi)  +8 G
  \left[ I_G({\cal M}) I( m^2_\pi) +\left(2{\cal M}^2-
  \frac{m^2_\pi}{4}\right)  I^2(m^2_\pi)\right] \right\}\;,
\label{mpiOPT}
\end{equation} 
\noindent
generalizing a similar expression in the large-$N_c$ limit (given here by the
first ${\cal O}(G)$ term in the right-hand side of Eq. (\ref{mpiOPT}), and for
$M^{\rm OPT}_q\to M^{\rm LN}_q$). We will use Eq. (\ref{mpiOPT}) to derive
some of the NJL parameters consistently at OPT level  by matching it to the
pion mass's experimental value.

\subsection{Pion decay constant}

 In view of the more convenient generalization at two-loop order,
we define the pion decay constant as  the axial-vector to
axial-vector current vacuum-to-vacuum transition\footnote{This definition is equivalent 
  to the more standard NJL one in terms of the one-pion to vacuum
  transition~\cite{klevansky},  provided that one uses a covariant-preserving
  regularization, and avoids at two-loop order the rather involved calculation of the
  pion-quarks coupling $g_{\pi q q}$.}:  

\begin{equation} 
\langle 0| T A^i_\mu(q) A^j_\nu(0)|\rangle = i g_{\mu\nu} \delta^{ij} f^2_\pi
+ {\cal O}(q_\mu q_\nu) \;,
\label{fpidef}
\end{equation} 
\noindent
where   $A^i_\mu \equiv \bar \psi \gamma_\mu \gamma_5 (\tau^i/2) \psi$.  At one-loop
order we recover the well-known expression~\cite{klevansky} 

\begin{equation} 
f^2_\pi(\mbox{1-loop}) = -4i N_c m^2 I(0) \;,
\label{fpi1loop}
\end{equation} 
\noindent
where $m$ will be the appropriate expression for the mass gap depending on the
approximation used, so that in our OPT case we have $m\to {\cal M}$,  as
defined in Eq. (\ref{calM}).   At two-loop we have a  vertex-type correction
diagram similar to the first diagram shown in {}Fig. \ref{fig2loop},   but
with the replacement: $ i \gamma_5 \tau_i \to i \gamma_5 \gamma_\mu
\tau_i/2$. The calculation is given in more detail in Appendix B, and we
obtain the simple result: 

\begin{equation}
f^2_\pi(\mbox{2-loop,vertex}) =   8 G N_c (n_\pi-1)  m^4 I^2(0) \;.
\label{fpi2loop}
\end{equation}   
\noindent
We thus can write a final expression for $f^2_\pi$, including the mass
insertion that contains consistently the other two-loop diagrams shown in
{}Fig. \ref{fig2loop}, as well as higher-order OPT corrections: 

\begin{equation} 
f^2_\pi = -4i N_c {\cal M}^2 I({\cal M} ) +8 (n_\pi-1) G N_c \: {\cal M}^4
I^2({\cal M}) \;.
\label{fpi2OPT}
\end{equation}
 
\subsection{Fitting the basic parameters in vacuum}

We now discuss the determination of the relevant basic parameters consistently
for the OPT analysis. The NJL parameter fit procedure is well known (see
e.g. \cite{buballa}) and adapted to  our generalized OPT expressions as we
discuss now.  More precisely, the OPT  expression for $f_\pi$, see
Eq. (\ref{fpi2OPT}), is fitted to the experimental value $f_\pi \sim 92.4$ MeV
together with the gap-equation Eq. (\ref{optMassVac}) used  to determine the
mass gap $M^{\rm OPT}_q$ and the coupling $G$. The actual fit is performed
using the sharp-cutoff regularization to be consistent with latter $T,\mu$
dependent quantities, so that we consider the noncovariant cutoff version of
Eq. (\ref{fpi2OPT}), which reads (from now on we take $N_{\rm f}=2$): 

\begin{equation} 
f^2_\pi = 4 N_c\: {\cal M}^2 I_4(0) -16 G N_c\:{\cal M}^4 I^2_4(0)\;,
\label{fpifinal}
\end{equation} 
\noindent
where ${\cal M}$ is defined in Eq. (\ref{calM}) and we have defined the
cutoff-dependent 3D integral equivalent to $I(q^2)$ in Eq. (\ref{defI}) as 

\begin{equation}
I_4(q^2) = \frac{1}{8\pi^2} \left[ \ln
  \left(\frac{\Lambda+\sqrt{\Lambda^2+m^2}}{m}\right)  
-\sqrt{4\frac{m^2}{q^2}-1} \:
  \tan^{-1}
  \left(\frac{\Lambda}{\sqrt{\Lambda^2+m^2}\sqrt{4\frac{m^2}{q^2}-1}}\right)
  \right]\;,
\label{I4div}
\end{equation} 
\noindent
similarly to Eqs. (\ref{I1div})-(\ref{I2div}) (and of course 
$m={\cal M}$ at OPT
order).  {}For completeness, we also give its analytical expression for
$q^2\to 0$, relevant for Eq. (\ref{fpifinal}): 

\begin{equation} 
I_4(0) = \frac{1}{8\pi^2} \left[ \sinh^{-1}\left (\frac{\Lambda}{m}\right
  )-\frac{\Lambda}{\sqrt{\Lambda^2 + m^2}} \right]\;.  
\end{equation} 
\noindent
In addition, similarly the 3D cutoff version of expression (\ref{mpiOPT}) reads

\begin{equation} 
\frac{m_c}{M^{\rm OPT}_q} = 4 G N_f N_c m^2_\pi \left\{ I_4(m^2_\pi)  +8 G
  \left[ I_2(0,0) I_4(m^2_\pi)/2 -\left(2{\cal M}^2- \frac{m^2_\pi}{4}\right)
  I^2_4(m^2_\pi)\right] \right\}\;,
\label{mpifinal}
\end{equation}  
\noindent
with $I_2$ defined in Eq. (\ref{I2div}),  which is fitted to the experimental
pion mass $m_\pi \sim 135$ MeV  to determine the bare (current) mass $m_c$ and
the coupling $G$ respectively, as functions of the other parameters. Then, we
can either consider the cutoff $\Lambda$ as an input parameter and derive all
quantities ($G$, $m_c$ and $M_q$,  $\langle \bar q q \rangle$) as functions of
$\Lambda$, or alternatively, determine $\Lambda$ for a given $\qq$ input
value.

\begin{figure}[tbh]
\epsfig{figure=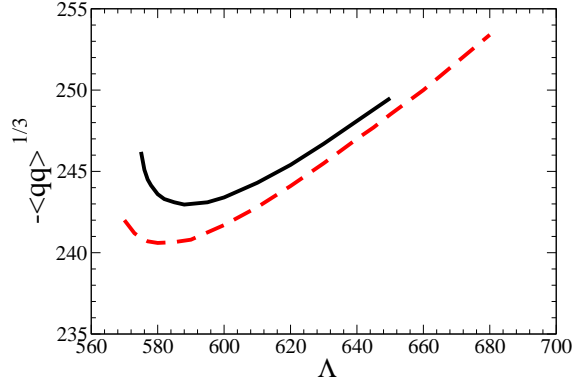,angle=-90,width=8cm}
\caption{(color online) Quark condensate as function of $\Lambda$ for $T=0$.   The OPT is
  represented by the continuous line and the LN approximation by the dashed
  line. Both quantities are in MeV units.}
\label{qqfig}
\end{figure}

\begin{figure}[tbh]
\epsfig{figure=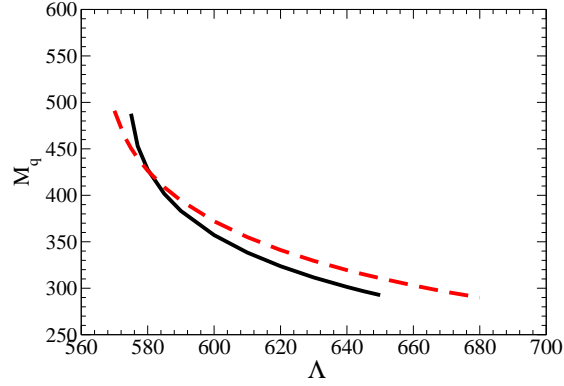,angle=-90,width=8cm}
\caption{(color online) Effective quark mass gap as a function of $\Lambda$ for $T=0$.  The
  OPT is represented by the continuous line and the LN approximation by the
  dashed line. Both quantities are in MeV units. }
\label{Mfig}
\end{figure}

Determination of $\Lambda$ for a given $\qq$ input in addition to $G$ and $m_c$ can
be done, for example, by inverting  the gap equation, solved for $\Lambda$, and using
the relation between the quark condensate and the mass gap:

\begin{equation}
 \qq = -\frac{M^{\rm OPT}_q-m_c}{4G}\;,
\label{qqopt}
\end{equation}
\noindent
noting that the last equation remains unmodifed with respect to the LN case
(except for the obvious replacement $M^{\rm LN}\to M^{\rm OPT}$).  
In both
cases, $\qq$ values are varied within a certain  range of $\Lambda$ input
values. Results are summarized in {}Figs. \ref{qqfig} and \ref{Mfig}, for the
quark condensate and mass gap, respectively, with particular values given in
Table~\ref{tabfit}.  The resulting values are roughly consistent  with the
allowed range from precise  $\qq$ determinations, in particular recent ones
from lattice calculations~\cite{qqlatt} or spectral sum
rules~\cite{qqSR}. Although the lattice determinations in particular have
considerably restricted the allowed $\qq$ value recently, they still allow
rather conservative range of variation of $\qq$ because of the relatively large
systematic uncertainties. 

As one can see, the mass gap and the quark
condensate are quite sensitive to the value of the cutoff, especially for the
mass gap, as expected, when it approaches a range where $M_q/\Lambda$ is no
longer small. Note that there  exist minimal values of $\Lambda$ (or
alternatively minimal values of $-\qq$) to obtain a self-consistent solution
to all relevant quantities. This is similar in the LN case, except that those
``theoretical'' lower and upper bounds are somewhat more restrictive in the
OPT case.  The lower bounds are, in the OPT case, approximately $\Lambda \sim
575$ MeV, and $-\qq \sim (242 {\rm MeV})^3$, as is clear from the figures,
thus producing the considered  range of $\Lambda$ and $\qq$. The LN
approximation allows a  rather similar   lower value of $-\qq \sim (240 {\rm
  MeV})^3$, but for a lower minimal  value of $\Lambda$~\cite{buballa}.
Therefore, the consistency with the OPT expressions restricts more the
possible $\qq$ range than in the LN case, since we only obtain solutions in
the range $242 {\rm MeV} \lsim -\qq^{1/3} \lsim 250$ MeV.  Note also that, for
relatively low values of $\qq $,  there are twofold solutions of $\Lambda$ for
$\qq$ input, as is clear from the figure: This is inherent to the structure of
the determining equations. It is easy to get rid of one of the branch solutions
by excluding unacceptably large values of $M_q$ (which on one of the branches,
not shown on Fig. \ref{Mfig}, grows very rapidly as a function of  $\Lambda$,
with $M_q >500$ MeV). Thus we only consider the branch with reasonable  values
of the constituent quark mass, as shown in {}Fig. \ref{Mfig}.
\begin{table}[htb]
\begin{center}
\caption{\label{tabfit} Basic parameters $G$, $m_c$ values and comparison of
  $M_q$, $\qq$  predictions from fitting  input data $m_\pi = 135$ MeV; $f_\pi
  = 92.4$ MeV: a) as function of cutoff $\Lambda$; b) fitting $\Lambda$ from
  $\qq$. All mass parameters are in MeV units and the bag constant, ${\cal
    B}$, is given in ${\rm MeV}/{\rm fm}^{3}$.} 
\begin{tabular}{l||l|l||l|l|l|l|l}
\hline \hline $\Lambda$ input [MeV] &  $G \Lambda^2$ & $ m_c $& $M_q$ &
$-\qq^{1/3}  $ &  $\frac{-2 m_c \qq}{f^2_\pi m^2_\pi }$ &${\rm Re}[m_\sigma]$
& ${\cal B}$ \\ \hline\hline (OPT-I)       580   &  2.46 &  5.0     & 427.7 &
243.6 & 0.93 &911.5&199.11 \\ \hline (LN-I)       580      &  2.54    &   5.6     &
426.5 & 240.6 & 1.001 &  856&162.58  \\ \hline   (OPT-II) 640          &  1.99    &
4.9 & 301.4 &  248  & 0.95 &  669 &91.00 \\ \hline (LN-II)   640       &  2.14    &
5.2 & 319.5  &  247    & 1.00 &  644&87.90  \\ \hline \hline $-\qq^{1/3}$ input & $G
\Lambda^2$ & $m_c$ & $M_q$ & $\Lambda$ &$\frac{-2 m_c \qq}{f^2_\pi m^2_\pi }$
& $Re[m_\sigma]$ &${\cal B}$\\ \hline \hline (OPT-III) 250 &  1.95 & 4.8 & 300  & 653  &
0.96 &  668&84.45  \\ \hline  (LN-III)  250 &  2.08 & 5.0 & 303.5  & 659 &1.00 & 680&78.88
\\ \hline  
\end{tabular}
\end{center}
\end{table}
One may remark at this stage that the corrections brought about by the OPT to
the mass gap and quark condensate, for a given $\Lambda$, are rather moderate
for the mass and even smaller for $\qq$. These OPT corrections to LN can be a
priori of any sign, depending on $\Lambda$ values, but with  a tendency for
the OPT mass gap  to be slightly lower than the corresponding LN one, with a
maximal departure of about $\sim 20$ MeV for $\Lambda \sim 630-640$
MeV. Indeed, it is worth mentioning that the bulk of OPT correction as compared with the LN
result is coming from the very first term in the right-hand side of
Eq. (\ref{fpifinal}). In comparison,  
the ${\cal O}(G)$ two-loop correction gives a much more
moderate effect, suppressed by a relative ${\cal M}^2/\Lambda^2$, making 
about 1 \% of the total contribution to $f_\pi$ e.g. for parameter set II of Table \ref{tabfit}. 
In fact, since  ${\cal M}\simeq [1+1/(4N_c)] M^{\rm
  OPT}_q\simeq 1.08 M^{\rm OPT}_q$, from Eq. (\ref{calM}),  as induced by the OPT ``nonperturbative''
corrections from Eq. (\ref{GAPPMS}), and replaces  $M^{\rm LN}_q$ at first order in
Eq. (\ref{fpifinal}), one may have $M^{\rm OPT}_q < M^{\rm LN}_q$ from fitting
the precise value of $f_\pi$ with Eq. (\ref{fpifinal}).

Then, from the OPT expressions of $f_\pi$, $m_\pi$ and $\qq$, one quantity of
interest which is  straightforward to determine within the OPT is the
Gell--Mann-Oakes-Renner (GMOR) relation~\cite{gmor}:  The latter can be
conveniently defined as  

\begin{equation} 
R_{\rm GMOR} \equiv -2m_c \frac{\qq}{f^2_\pi\,m^2_\pi} \;,
\end{equation} 
where $R^{\rm LN}_{\rm GMOR}\simeq 1$ in LN, up to tiny ${\cal O}(m^2_c/M^2)$
corrections.  The ``exact'' values of the OPT corrections $R^{\rm OPT}_{\rm GMOR}$  for
a given $\Lambda$ are immediately obtained from the fitted values of $m_c$, and
$\qq$ and are given in Table \ref{tabfit}. It may be useful, in addition,  to
give an approximate expression by combining the expressions for $f_\pi$,
$m_\pi$, and $\qq$, given, respectively, in Eqs. (\ref{fpifinal}),
(\ref{mpifinal}) and (\ref{qqopt}):   Expanding those relations to first order
in $G$, neglecting the small $q^2$ dependence inside the $I_4(q^2)$ integral
(i.e. assuming $I_4(m^2_\pi)\simeq I_4(0)$), and finally neglecting the tiny
${\cal O}(m^2_c/M^2)$ corrections, we obtain: 

\begin{equation} 
R^{\rm OPT}_{\rm GMOR} \simeq \left (1+\frac{1}{4N_c} \right)^{-2}\: \left\{
  1+4\,G\left[I_2(0,0)-5{\cal M}^2 I_4(0)+\frac{m^2_\pi}{2}
  I_4(0)\right]\right\}\;.
\label{GOR}
\end{equation}    
Again, we remark that a large part of the correction with respect to LN comes
about simply from the factor $(1+1/(4N_c))^{-2}\sim 0.85$  induced by the
specific nonperturbative OPT relation in Eq. (\ref{calM}), although this is partly compensated
by the (positive) ${\cal O}(G)$ corrections  (formally of order $1/N_c$ since $G\sim 1/N_c$) 
inside the bracket of Eq. (\ref{GOR}) (which is about $10\%$  for the relevant values of the
parameters), so that the final OPT (negative) corrections in Table
\ref{tabfit} are not more than a few percent.\\  It is instructive to
  compare at this stage those results with the present theoretical status and
  constraints on the GMOR relation. As is well-known, the latter corresponds
  to the leading term in the expansion in powers of the quark masses.
  Theoretically, Chiral Perturbation Theory (ChPT) \cite{chpt} typically
  predicts, at next-to-leading orders \cite{leutwyler1}, a few percent decrease from
  $R^{\rm GMOR}=1$, and  even somewhat larger deviations have been advocated \cite{descotes1}
  in a generalized ChPT framework (where contributions of different ChPT orders could compete).  
  Experimentally, one can extract indirectly the GMOR relation (or equivalently the value  of the
  relevant ChPT parameters) from measurements of the $\pi\pi$ $S$-wave scattering
  lengths in $K\to \pi \pi l \nu$ scattering~\cite{pipidata}, upon additional
  theoretical assumptions and experimental information (see e.g. 
  \cite{descotes1,descotes2} for a discussion). The latest precise measurement of the
  relevant S-wave scattering lengths \cite{NA48},  together with the most
  recent analysis performed in the  framework of two-loop chiral perturbation
  theory,  are consistent with a value $R^{\rm GMOR} \sim
  0.94$ \cite{leutwyler1,leutwyler2} with a few percent accuracy. Thus, although considering the simplest
  $SU(2)$-symmetric NJL model can hardly compete with the latest sophistication level of ChPT
  to describe a fully realistic
  phenomenology, it is quite satisfactory at least that the OPT corrections we
  obtain appears roughly consistent with present constraints.

Another quantity of interest, at $T=0$ and $\mu=0$, is the ``bag constant'',
defined in terms of the pressure as~\cite{buballa}
\begin{equation}
 {\cal B} = P(M_q) - P(m_c) \,\,.
\end{equation}
As shown in Tab. \ref{tabfit}, this quantity is very sensitive to the
parameter set used and ranges from $84.45 {\rm MeV}/{\rm fm}^3$ to  $199.11
{\rm MeV}/{\rm fm}^3$ within the OPT and from $78.88 {\rm MeV}/{\rm fm}^3$ to
$162.58 {\rm MeV}/{\rm fm}^3$ within the LN approximation

{}Finally, for completeness we evaluate the OPT expression for the $\sigma$
meson mass  (at $T=\mu=0$) as usual by a procedure very similar to the one for
the pion mass above, where the two-point function is that for a scalar, i.e.,
with $i \gamma_5 \tau_i$ replaced by $ 1$ in both flavor and Dirac spaces
within, e.g., Eq.  (\ref{pionv2l}) in Appendix B.  At the one-loop level, the
geometric resummation of this diagram produces a pole at $q^2 = 4 M_q^2$ in
the chiral limit, where $M_q$ is the mass gap. Evaluating all
quantities including the OPT corrections induced at two-loop, and using the
gap equation consistently at this order, gives a correction of ${\cal O}(G)$
to the well-known LN relation, and we are led  to the final scalar meson mass
given as the solution of the  implicit equation    

\begin{eqnarray} 
m^2_\sigma &=& 4 (M^{\rm OPT}_q)^2 +\frac{m_c}{M^{\rm OPT}_q}\left(4 G N_{\rm f} N_c
I(m^2_\sigma)\right)^{-1} \nonumber \\ &+&\frac{1}{4G\,I(m^2_\sigma)}\left[i
  +8G\,I_G({\cal M}) -i\:\sqrt{1-16iG \,I_G({\cal M})  +64G^2\,I^2_G({\cal
      M})} \right]\;,
\label{msigma}
\end{eqnarray}  
with $I(q^2)$ as defined in Eq. (\ref{defI}). Note that actually, the OPT
solution gives a relatively small correction to the one-loop relation
$m^2_\sigma =4 m^2 +{\cal O}(m_c)$, as can be seen more clearly by expansion of
Eq. (\ref{msigma}) to first order in $G$: 

\begin{equation} 
m^2_\sigma = 4 (M^{\rm OPT}_q)^2 +\frac{m_c}{M^{\rm OPT}_q}\left[4 G N_{\rm f} N_c
I(m^2_\sigma)\right]^{-1} -16 i G \frac{I^2_G({\cal M})}{I(m^2_\sigma)} +{\cal
  O}(G^2)\;.  
\end{equation}  
Numerically, for the set II input values in Table \ref{tabfit}, corresponding
to $\Lambda=640$ MeV, we obtain: $m^{\rm OPT}_\sigma \simeq 669 +17\,i$ MeV,
to be compared with the corresponding LN value for the same input: $m^{\rm
  LN}_\sigma \simeq 644 +0.8\,i$ MeV. Thus it gives a few percent correction
to the standard (i.e LN) NJL result. Other values are given for
illustration\footnote{Note that the presence of the imaginary parts, as usual,
  reflects that $m_\sigma$ moves above the $q\bar q$ threshold which gives an
  imaginary part to $I(q^2)$, which is related to the fact that the NJL model
  cannot accommodate the quark confinement.} in Table \ref{tabfit}.  
  Thus the standard NJL picture at leading LN order predicting a sharp resonance,   
  conflicting the very large $\sigma$ width, is not drastically 
  modified by OPT first order corrections which basically use the same (variationally modified) NJL
  Lagrangian. Of course it would be possible in principle within OPT framework to evaluate the
  dominant $\sigma \to \pi\pi$ decay mode by calculating an  
  effective $\sigma \pi \pi$ coupling, via a quark loop vertex graph
  with external $\pi$ and $\sigma$, thus providing some definite corrections to such similar LN approaches 
  (which roughly gave the right
  order of magnitude of the $\sigma$ width (see e.g. \cite{njlrev3}
  and ref. therein)). However, already at LN order this raises a number of conceptual problems,
  and anyway such OPT corrections would certainly not accomodate all
  phenomenologically realistic properties of the $\sigma$ meson.  As
  is well-known  the $\sigma$ resonance had a long history with many 
  controversies on its status, which is not yet fully  clarified \cite{mesonrev}.
  We therefore refrain to investigate more details on the $\sigma$ meson mass
  and other properties, which is well beyond the scope of the present work.

Before considering the model for non-zero temperature and chemical
  potential, it may be worth  to review, at this stage, some definite differences
  betwen the OPT corrections obtained for the vacuum quantities and some other  approaches
  beyond mean-field approximation, like typically the corrections obtained
  from the  $1/N_c$ expansion \cite{nloNc,oertel}. Note that our OPT corrections
  beyond the LN/MFA results for all quantities ($M^{\rm OPT}$, $f_\pi$, $\qq$, ...) 
  are formally organized to be of order ${\cal O}(1/N_c$, since $G\sim 1/N_c$ in a consistent
  $1/N_c$-expansion framework. However, as discussed in introduction, OPT differs in several 
  respect from the genuine $1/N_c$ expansion, both qualitatively and quantitatively. 
  More precisely, some relevant differences are as follows:
\begin{itemize}
 \item As we already mentioned, at first OPT order the relevant two-loop
   calculations actually reduce to one-loop squared contributions, 
   thus regularized by a single standard cutoff $\Lambda$, in contrast with the genuine $1/N_c$
   corrections where an independent cutoff $\Lambda_M$ parameter is sometimes 
   introduced to regularize the (meson) loops~\cite{oertel}.
\item The mechanism satisfying the Goldstone theorem in OPT, via
  (perturbative) cancellations of  different contributions, is somewhat
  similar to the one at work in the $1/N_c$  case \cite{nloNc}, except that the
  latter involves extra $1/N_c$ contributions that are higher OPT
   orders (i.e. ${\cal O}(\delta^2 G^3)$), and therefore omitted in the 
OPT case.  
  (Note that the consistency of OPT with the Goldstone theorem was also 
  recently shown in the different context of the $O(N)$ $\phi^4$ model~\cite{chhat}.)  
\item in the $1/N_c$ framework one obtains a correction to the standard NJL
  (LN) $M_q$ versus $\qq$ relation, Eq. (\ref{qqopt}), while the latter LN
  form is preserved at first OPT order (with   simply the replacement $
  M_q^{\rm LN}\to M_q^{\rm OPT}$). This has another practical consequence concerning
  the  GMOR relation, which receives only higher $1/N^2_c$ order corrections
  within the $1/N_c$ framework \cite{oertel},   while OPT gives formally 
  explicitly $1/N_c$ corrections, in Eq. (\ref{GOR}) (though those are partially 
  canceled by  the extra ${\cal O}(G)\sim 1/N_c$ terms, as already  discussed above).
\end{itemize}

We will consider each of the relevant cases as a function of the
temperature and chemical potential. To compare OPT versus standard LN results,
we will  take as a representative case for input parameter values mainly those
from set II from Table \ref{tabfit}, corresponding to $\Lambda=640$ MeV, which
gives reasonable values both for the quark mass and condensate. We have also
studied the dependence of our main results upon varyiations of
those input parameters
within acceptable ranges, and will comment on this dependence when it is
relevant.

\section{Numerical results at finite temperature and density}

Let us now turn to the study of the effects of temperature and chemical
potential in the NJL model within the OPT starting with the finite-temperature and zero-chemical
potential case.

\subsection{Hot matter at zero density}

\begin{figure}[tbh]
\vspace{0.5cm} \epsfig{figure=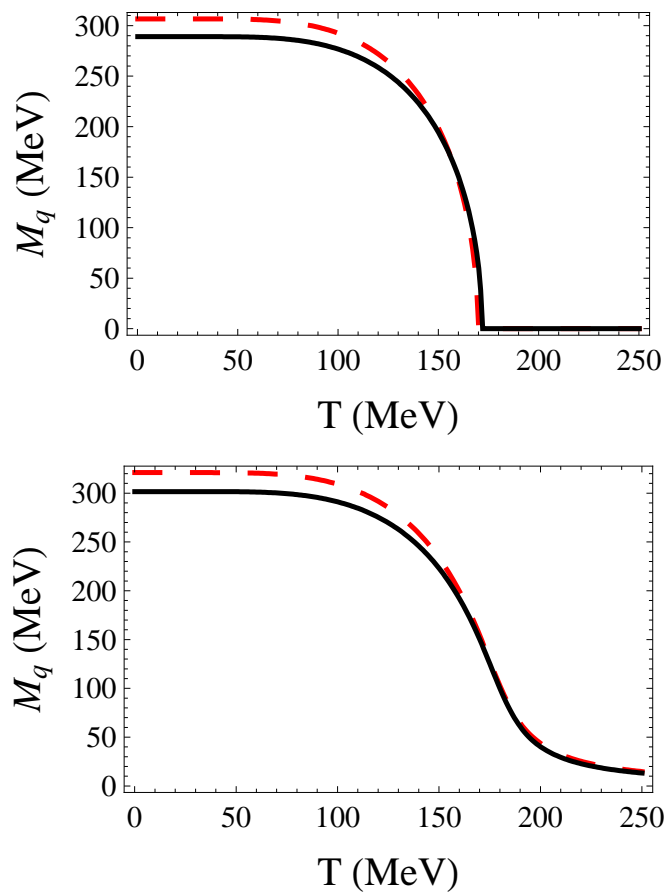,angle=0,width=10cm}
\caption{(color online) Constituent quark mass  as a function of $T$ for
  $\mu=0$ with $m_c=0$ (top) and  for the case of nonzero $m_c$ (bottom):
  $m_c=4.9 \rm MeV$ (OPT) and $m_c=5.2 \rm MeV$ (LN). The continuous line
  represents the OPT result and the dashed line represents the LN approximation.}
\label{MqT}
\end{figure}

This scenario is important for observation the melting of the condensate, ${\bar
  \sigma}_c$, owing to the appearance of thermal effects.  Theoretically, this
case can be more easily exploited by use of lattice techniques because of 
the absence
of the  sign problem in the partition function at zero chemical potential.
Since $\mu=0$, Eq. (\ref{GAPPMS})  still holds and the thermal mass is
readily obtained from the solution of the gap equation,

\begin{eqnarray} 
M^{\rm OPT}_q(T)&=& m_c+{\bar \sigma}_c^{\rm OPT}= m_c + 4 G N_{\rm f}N_c \;
{\cal M}  I_2(\mu=0,T\neq 0)\Bigr|_{\bar{\eta}+m_c = {\cal M}} \,,
\label{gapM}
\end{eqnarray}
which is solved numerically. The OPT results for Eq. (\ref{gapM}) are compared
with the  LN results in {}Fig. \ref{MqT} for the case of finite and vanishing
current mass respectively.  The OPT displays a second order phase transition
taking  place at $T_c\simeq 172 \,  {\rm MeV}$, while the LN result is just a
bit smaller at $T_c\simeq 170 \, {\rm MeV}$. 
It is useful to compare these results with recent next to leading order (NLO)
corrections to the LN approximation obtained within the 2PI (or
$\Phi$-derivable) functional formalism~\cite{newbuballa}. It is shown
in that case that the NLO corrections decrease the critical temperature in the
chiral case. This is in a sense expected, due to the effect of the inclusion 
of fluctuations. In the OPT the results depart little from
the LN ones at finite temperature and zero density at this lowest order. However, as we will be
seen below, the inclusion of finite density effects result in much more
pronunced change of the critical quantities and overall behavior of the
thermodynamical quantities as compared to the LN case.
Also seen e.g. from {}Fig. \ref{MqT}, the OPT predicts a slight
smaller crossover temperature as compared with the LN case, which is in
agreement with the recent results obtained in \cite{newbuballa} from NLO
corrections obtained with the 2PI formalism.

All basic thermodynamic quantities can be derived from the
pressure $P$, obtained 
from the free energy density
Eq.~(\ref{landauenergy}) using  $P=-{\cal F}(\bar{\sigma})$, with $\bar{\sigma}$
obtained from the gap equation~(\ref{simpleGAP}). It is also 
convenient to work directly in terms of the normalized pressure, 
$P_N(T,\mu) = P(T,\mu)-P(0,0)$ so that the
energy density also vanishes at $T=0$ and $\mu=0$ (note that, although this subtraction 
at $T=0$ and $\mu=0$ is usual in the literature and it is done so to make contact with 
standard treatments in lattice calculations, it also removes some scale dependence). 
{}For simplicity we drop the subscript $N$ in all relations involving $P$ in
the following.

\begin{figure}[tbh]
\vspace{0.5cm} \epsfig{figure=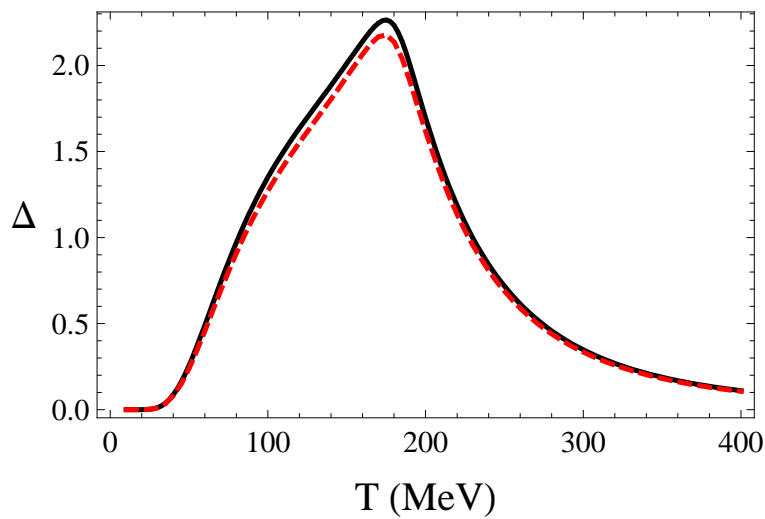,angle=0,width=10cm}
\caption{(color online) Interaction measure, $\Delta$, as a function of
  $T$ for $\mu=0$.  The continuous line represents the OPT result and the
  dashed  the LN approximation.}
\label{intmeas}
\end{figure}

\begin{figure}[tbh]
\vspace{0.5cm} \epsfig{figure=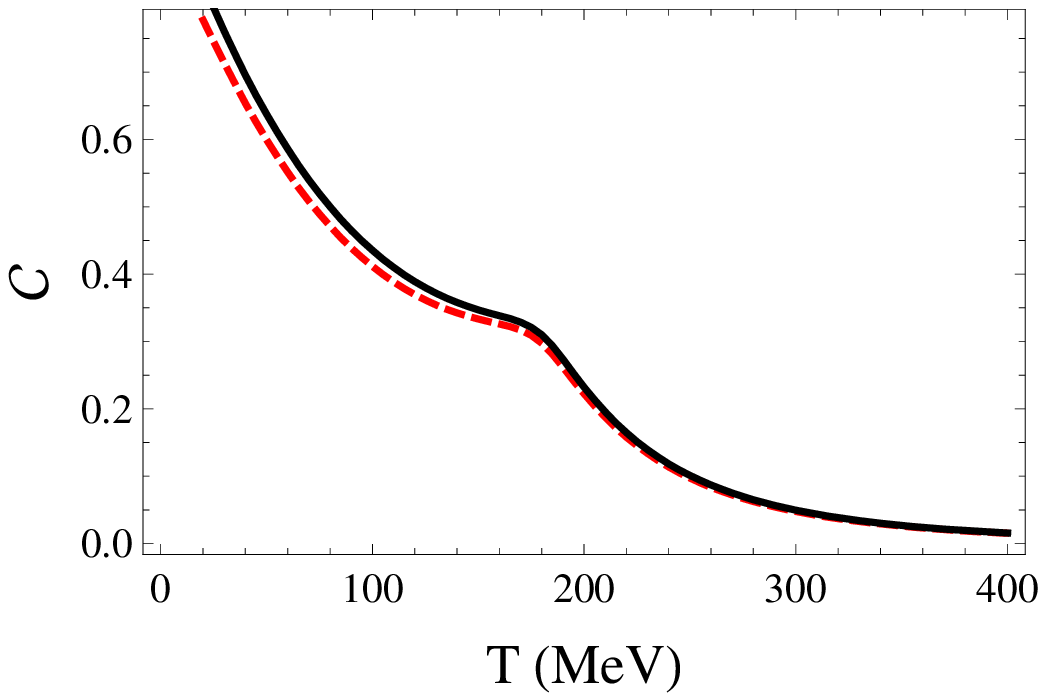,angle=0,width=10cm}
\caption{(color online) Conformal measure, ${\cal C}$, as a function of
  $T$ for $\mu=0$.  The continuous line represents the OPT result and the
  dashed  the LN approximation.}
\label{confmeas}
\end{figure}

\begin{figure}[tbh]
\vspace{0.5cm} \epsfig{figure=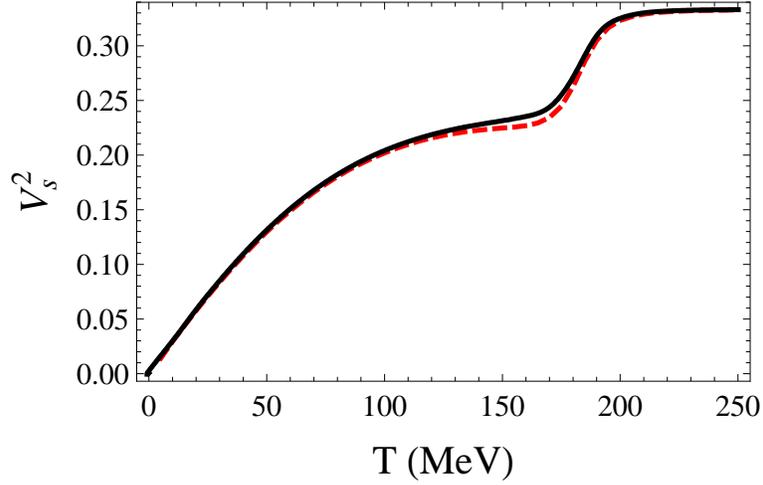,angle=0,width=10cm}
\caption{(color online) Sound velocity squared, $V_s^2$, as a function of
  $T$ for $\mu=0$.  The continuous line represents the OPT results and the
  dashed  the LN approximation.}
\label{sound}
\end{figure}

\begin{figure}[tbh]
\vspace{0.5cm} \epsfig{figure=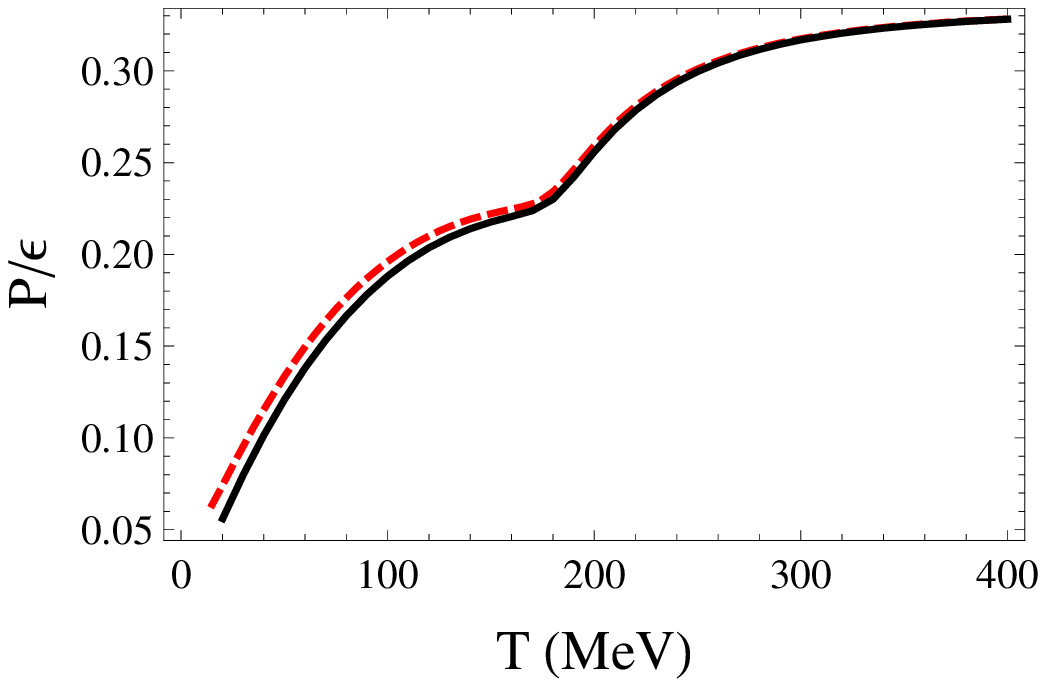,angle=0,width=10cm}
\caption{(color online) Equation of state parameter, $w=P/\epsilon$, as a
  function of $T$ for $\mu=0$.  The continuous line represents the OPT result
  and the dashed  the LN approximation.}
\label{eqst}
\end{figure}

In {}Figs. \ref{intmeas}, \ref{confmeas}, \ref{sound} and  \ref{eqst} we
give as a function of the temperature,  respectively, the interaction measure
(or trace anomaly) $\Delta$,  the conformal measure ${\cal C}$, the sound
velocity squared $V_s^2$, and the equation of state parameter $w$. These
quantities  are defined, as
usual, by the expressions,

\begin{equation}
\Delta = \frac{\epsilon - 3 P}{T^4}\;,
\label{Delta}
\end{equation}
\begin{equation}
{\cal C} = \frac{\Delta}{\epsilon}\simeq 1-3 V_s^2\;,
\label{calC}
\end{equation}
\begin{equation}
V_s^2 = \frac{dP}{d \epsilon}\;,
\label{Cs2}
\end{equation}

\begin{equation}
w=P/\epsilon\;.
\label{eqw}
\end{equation}
Here the energy
density, $\epsilon$, is defined as $\epsilon= -P +T\, s+\mu \, \rho$, where $s$ is the entropy density, $s=
(\partial P/\partial T)_\mu$, and $\rho$ the quark number density,
$\rho=(\partial P/\partial \mu)_T$.  The interaction measure is useful to give
the amount of scale violation through the different phases the system can
have, while the conformal  measure shows how the system might approach the ideal
gas case, in which ${\cal C}=0$, or $V_s^2=1/3$. Both quantities are expected
to peak near a phase transition or a crossover, thus being useful in locating
the critical line. Because of our choice of regularization for the thermal
integrals we expect that the OPT and LN will agree at high temperaures
reproducing the Stefan-Boltzmann limit.  This is shown in
{}Figs. \ref{intmeas} and  \ref{confmeas}.

{}Figures \ref{intmeas}, \ref{confmeas}, \ref{sound} and  \ref{eqst} show the
results for both the OPT and LN cases. We notice that the differences between
the two approximations are small for all these quantities, thus indicating a
robustness of the LN approximation when only thermal effects are
considered. As expected all quantities have a smooth behavior  around the
temperature values where the crossover takes place. The OPT predicts a
slightly higher  interaction measure and speed of sound at the crossover.

\subsection{Cold matter at finite density}

\begin{figure}[tbh]
\vspace{0.5cm}  \epsfig{figure=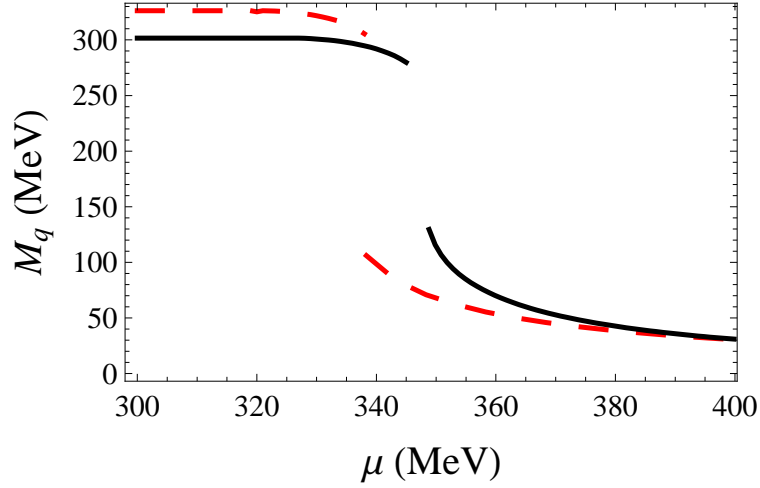,angle=0,width=10cm}
\caption{(color online) Effective quark mass, in units of the vacuum mass, as
  a function of $\mu$ for $T=0$. The OPT is represented by the continuous line
  and the LN approximation by the dashed line.}
\label{Mqmu}
\end{figure}

The effects caused by the OPT first-order corrections for cold and dense quark
matter can now also be studied at $T=0$ and $\mu \neq 0$.  Physically this
situation is relevant in studies related to neutron stars for example. This is
also the  case where lattice techniques face more problems with the sign
problem.   As discussed before, in this case the simple PMS gap relation,
Eq. (\ref{GAPPMS}), does not hold and we shall proceed numerically. Let us
first define the integrals $I_1$, $I_2$ and $I_3$ in the limit  $T \to 0$,
with $\mu \neq 0$. They read

\begin{equation}
I_0(\mu,T=0) =  \int \frac{d^{3}p}{\left( 2\pi \right) ^{3}}\theta(\mu-E_p)=
\frac{\theta(\mu-\eta-m_c)}{6 \pi^2}\left[\mu^2-(\eta+m_c)^2\right]^{3/2}\;,
\end{equation}

\begin{eqnarray}
I_1(\mu,T=0) - I_1(0,0) &=& \int \frac{d^{3}p}{\left( 2\pi \right) ^{3}}
(\mu-E_p) \theta(\mu-E_p) \nonumber \\ &=&\frac{\theta(\mu-\eta-m_c)}{32
  \pi^2}\left\{ (\eta+m_c)^{4}\ln \left[ \frac{ \left( \sqrt{\mu
      ^{2}-(\eta+m_c)^{2}}\mathbf{+}\mu \right) ^{2}}{(\eta+m_c)^{2}}\right]
\right. \nonumber \\ &+& \left. \frac{10}{3}\mu  \left[
\mu^{2}-(\eta+m_c)^{2}\right]^{\frac{3}{2}}{-2\mu ^{3}}\sqrt{{\mu
    ^{2}-(\eta+m_c)^{2}}}\right \}\;,
\end{eqnarray}
and

\begin{eqnarray}
I_2(\mu,T=0) - I_2(0,0)&=&   -\int \frac{d^{3}p}{\left( 2\pi \right) ^{3}}
\frac{1}{E_p} \theta(\mu-E_p) \nonumber \\ &=& - \frac{\theta(\mu-\eta-m_c)}{4
  \pi^2} \left \{\mu \sqrt{\mu
  {{^{2}}\mathbf{-}(\eta+m_c)^{2}}}-\frac{(\eta+m_c)^{2}}{2} {\ln }\left[
  \frac{\left( \mu \mathbf{+}\sqrt{\mu
      {{^{2}}\mathbf{-}(\eta+m_c)^{2}}}\right) ^{2}}{(\eta +m_c)^{2}}\right]
\right\}\;,
\end{eqnarray}
where $I_1(0,0)$ and $I_2(0,0)$ are given by Eqs. (\ref{I1div}) and
(\ref{I2div}), respectively.  Next, the mass is obtained by considering

\begin{equation} 
M^{\rm OPT}_q(\mu)= m_c + 4 G N_{\rm f}N_c (\bar {\eta}+m_c) I_2(\mu,T=0)\,\,,
\end{equation}
where the optimum, ${\bar {\eta}}$ is determined  numerically upon solving the
implicit equation,

\begin{equation}
 \left \{ \left[\eta -  4 G (\eta+m_c) {\cal G}(N) I_2 \right] \left[ 1 +
   (\eta+m_c) \frac {d}{d \eta} \right] I_2 + 4 G I_3 \frac{d}{d \eta} I_3
 \right \}_{\eta = {\bar \eta}}=0 \,\,,
\label{simplePMSTZero}
\end{equation}
which is obtained by substitution of Eq. (\ref{simpleGAP}) in
Eq. (\ref{simplePMS}). Note that this relation is valid only when one is interested in
the minimum of the free energy, as for example when evaluating the
pressure. Otherwise, Eq. (\ref{simplePMS}), has to be used.
{}Figure~\ref{Mqmu} shows the quark effective mass as a function of  $\mu$,
obtained with the OPT and the LN. The qualitative behavior is the same and the
critical values for the first order phase transition are   $\mu_c \simeq
348.83 \, {\rm MeV}$ for the OPT and $\mu_c \simeq 338.32 \, {\rm MeV}$ for
the LN approximation. 

\begin{figure}[tbh]
\vspace{0.5cm}  \epsfig{figure=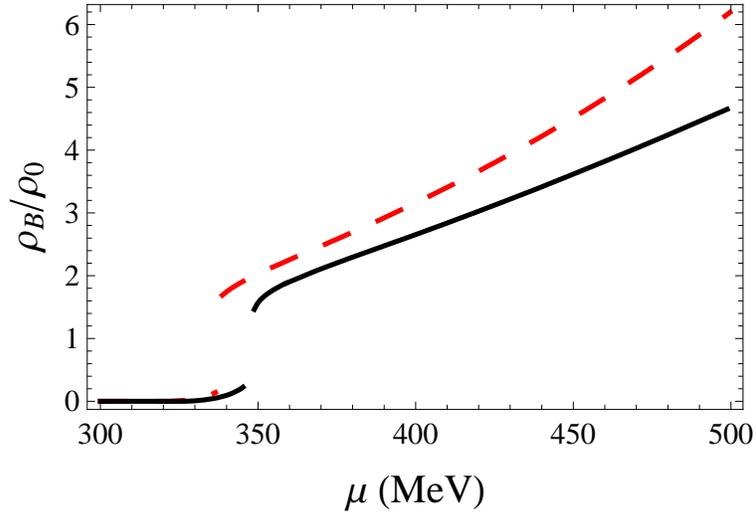,angle=0,width=10cm}
\caption{(color online) Baryonic density, in units of $\rho_0=0.17
  \, {\rm fm}^{-3}$, as a function of $\mu$ for $T=0$. The continuous line
  corresponds to OPT and the dashed line to the  LN approximation.}
\label{rhomu}
\end{figure}

\begin{figure}[tbh]
\vspace{0.5cm}   \epsfig{figure=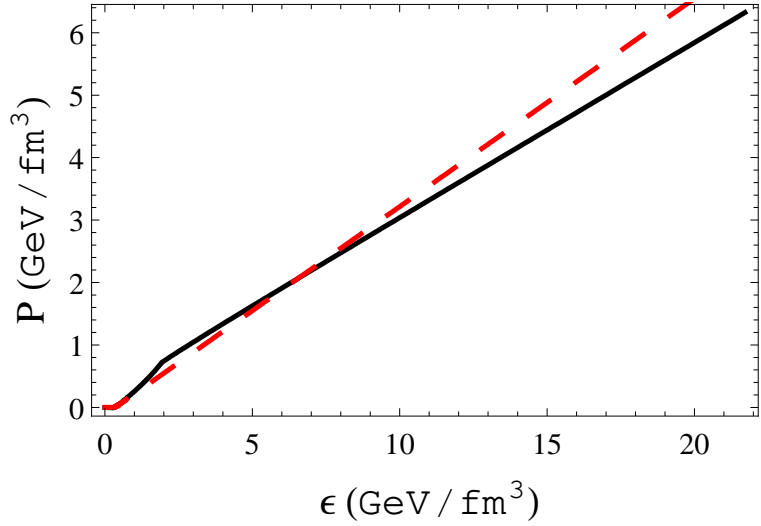,angle=0,width=10cm}
\caption{(color online) Equation of state  generated by the OPT (continuous
  line) and the LN approximation (dashed line).}
\label{Pepsilon}
\end{figure}

Once the free energy is evaluated and its minimum is located, one can 
obtain the pressure using $P=-{\cal F}(\bar \sigma)$. The density $\rho$ can
now be easily computed by using $dP/d\mu$ and taking into account the gap and
PMS equations.  This gives

\begin{equation}
\rho= 2N_f N_c I_1^\prime - 2 N_fN_c(m_c+{\bar \eta})({\bar \eta}-{\bar
  \sigma_c})I_2^\prime -  8GN_fN_cI_3 I_3^\prime + 4GN_fN_c(m_c+{\bar \eta})^2
I_2 I_2^\prime \,,
\end{equation}
where the primes indicate derivatives with respect to $\mu$.
{}Figure~\ref{rhomu} shows the baryonic density  $\rho_B= \rho/3$, in units of
normal matter density ($\rho_0= 0.17 \, {\rm fm}^{-3}$), as a function of
$\mu$.  At $T=0$, the energy density is given as usual by $\epsilon=-P+\mu
\rho$ and allows us to obtain the EoS as shown in {}Fig. \ref{Pepsilon} where
the OPT appears to generate a softer relation due to a more abrupt change of
slope at $\mu_c$.

\begin{figure}[tbh]
\vspace{0.5cm}   \epsfig{figure=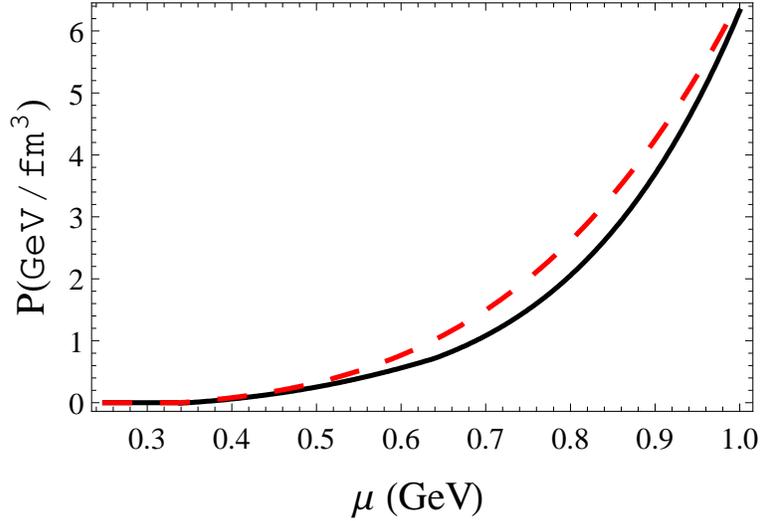,angle=0,width=10cm}
\caption{(color online) The pressure as a function of $\mu$ within the OPT
  (continuous line) and within the LN approximation (dashed line). }
\label{12}
\end{figure}

\begin{figure}[tbh]
\vspace{0.5cm}   \epsfig{figure=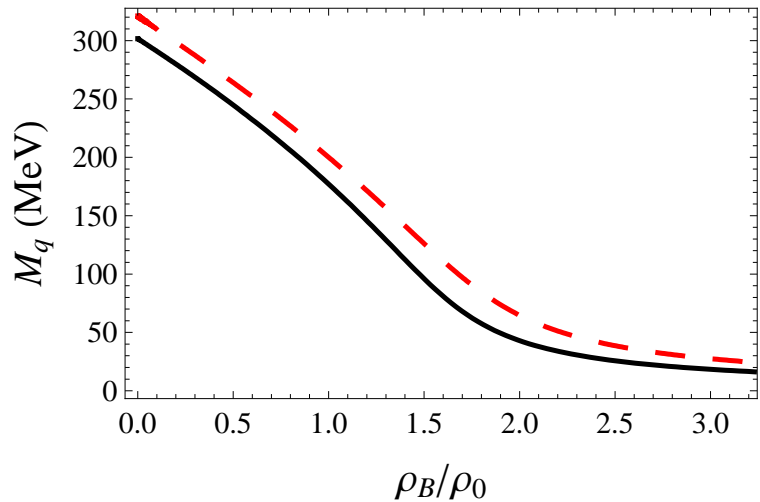,angle=0,width=10cm}
\caption{The OPT quark effective mass as a function of the density for
  $m_c=4.9$ MeV (continuous) and $m_c=0$ (dot-dashed).}
\label{massrho}
\end{figure}

\begin{figure}[tbh]
\vspace{0.5cm}   \epsfig{figure=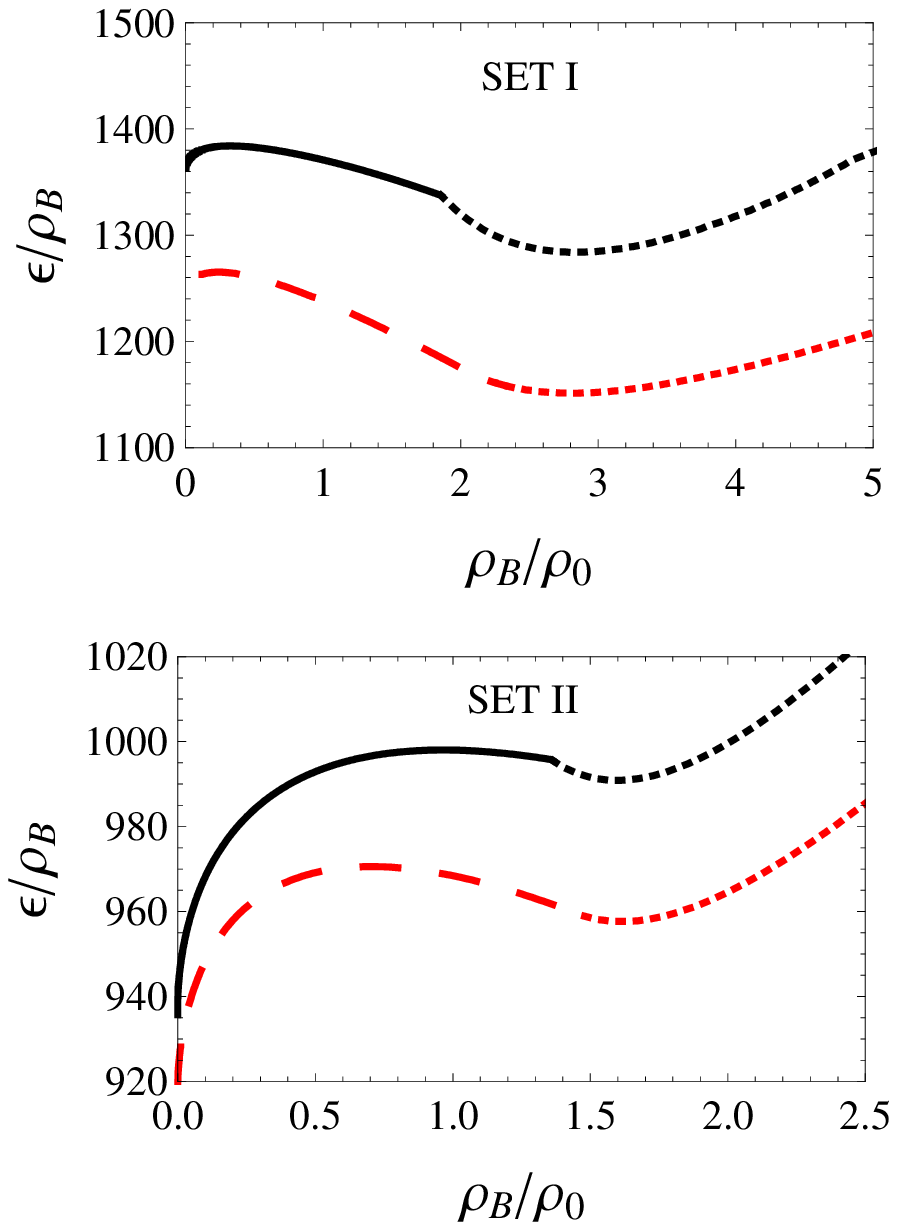,angle=0,width=10cm}
\caption{(color online) Energy per baryon number as a function of the baryonic
  density, in the chiral limit, for sets I and II as indicated in the figure.  For the massive solutions, the OPT result is represented by the continuous line and the  LN
  by the dashed line. The dotted lines represent the massless solutions in both cases. }
\label{Erho}
\end{figure}

We now look at the matter stability by considering the energy per baryon
number, $\epsilon / \rho_B=-P/ \rho_B+3\mu$, as a function of the baryonic
density, $\rho_B$. As Koch et al. \cite {koch} found out, the NJL does not
present a minimum at nuclear matter density, $\rho_0=0.17 \, {\rm fm}^{-3}$. The situation can be
remedied by introducing a vector coupling (see Refs
\cite{{buballa_stab},koch} for details).  However, in the minimal form of the
NJL considered in the present work one expects that stable matter can only
occur in the chiral restored phase \cite{buballa_stab}.  Plots of the OPT
quark effective mass in the chiral limit and away from it as functions of
$\rho_B$ are shown in {}Fig.  \ref{massrho}.    The energy per baryon number as
a function of the baryonic density is shown in {}Fig. \ref{Erho}, for both the
OPT and LN cases with two sets of parameters in the chiral limit to illustrate 
how the energetically favored point changes from $\rho_B=0$, within set II, to $\rho_B\simeq 2.5 \rho_0$, 
within set I, giving rise to stable matter.

Note that corrections beyond the LN approximation are more significant at
finite density and $T=0$ than vice versa, as shown from
Figs. \ref{Mqmu}-\ref{Erho}. This points to the importance of the effect of
fluctuations when studying dense matter. The same is true when  considering
the effects of both temperature and chemical potential, as we are going to
analyze in the next subsection. Before doing that let us see how the
critical temperature, $T_c (\mu=0)$, and the critical chemical potential,
$\mu_c (T=0)$, change with the different  sets of paramters. Here the quantity
$T_c (\mu=0)$ can be defined  only in the chiral limit
($m_c=0$) when a second order phase transition occurs for both approximations
and any set of parameters. Away from the chiral limit a cross over takes place
at $\mu=0$. On the other hand  $\mu_c (T=0)$ can be defined for $m_c=0$ and
$m_c \ne 0$ and in the first case a first order transition occurs for both
approximations and any set of parameters. However, when set III is considered
at $m_c \ne 0$ the LN predicts a very weak first order transition with a small
latent heat while the OPT predicts a cross over also at $T=0$ (and everywhere
in the $T-\mu$ plane). As emphasized in Ref. \cite {klevansky} this is a
possible scenario within the NJL where the order of the phase transition  is indeed 
very sensitive to input parameters. In this respect our sets I and II seem to
be more in lign with what one expects to happen in  more realistic
situations. Table \ref{tabfit2} summarizes the numerical results for these
critical quantities when the three parameter sets are employed by the two
approximations considered here.
\begin{table}[htb]
\begin{center}
\caption{\label{tabfit2} Critical temperature at zero chemical potential,
  $T_c$, and critical  chemical potential at zero temperature, $\mu_c$. Both
  quantities and the mass parameters are given in MeV. $T_c$ is evaluated at
  $m_c=0$ and $\mu_c$ at $m_c=0$ as well as at $m_c \ne 0$. For reference, the
  vacuum effective mass ($M_q$) evaluated at $m_c \ne 0$ is also shown.} 
\begin{tabular}{l||l|l||l|l|l|l}
\hline \hline $\Lambda$ input [MeV] &  $G \Lambda^2$ & $ m_c $& $M_q$ & $T_c
(m_c=0)  $ &  $\mu_c(m_c=0)$ &$\mu_c(m_c \ne 0)$  \\ \hline\hline (OPT-I)
580.0   &  2.5 &  5.0     & 427.7 &  198.0 & 428.0 &439.6 \\ \hline (LN-I)
580.0      &  2.5    &   5.6     &   426.5 & 190.0 & 383.7 &  396.2  \\ \hline
(OPT-II) 640.0          &  2.0    &  4.9 & 301.4 &  172.0  & 330.9 &  348.8
\\ \hline (LN-II)   640.0       &  2.1    &  5.2 & 319.5  &  170.0   & 319.4 &
338.3  \\ \hline \hline $-\qq^{1/3}$ input & $G \Lambda^2$ & $m_c$ & $M_q$ &
$T_c (m_c=0)  $ &  $\mu_c(m_c=0)$ &$\mu_c(m_c \ne 0)$ \\ \hline \hline
(OPT-III) 250.0 &  1.9 & 4.8 & 300.0  & 169.0 & 308.9 &  {\rm cross over}
\\ \hline  (LN-III)  250.0 &  2.1 & 5.0 & 303.5  & 167.0 &323.0 & 329.0
\\ \hline  
\end{tabular}
\end{center}
\end{table}

\subsection{The Phase Diagram and the Critical End Point}

\begin{figure}[tbh]
\vspace{0.5cm} \epsfig{figure=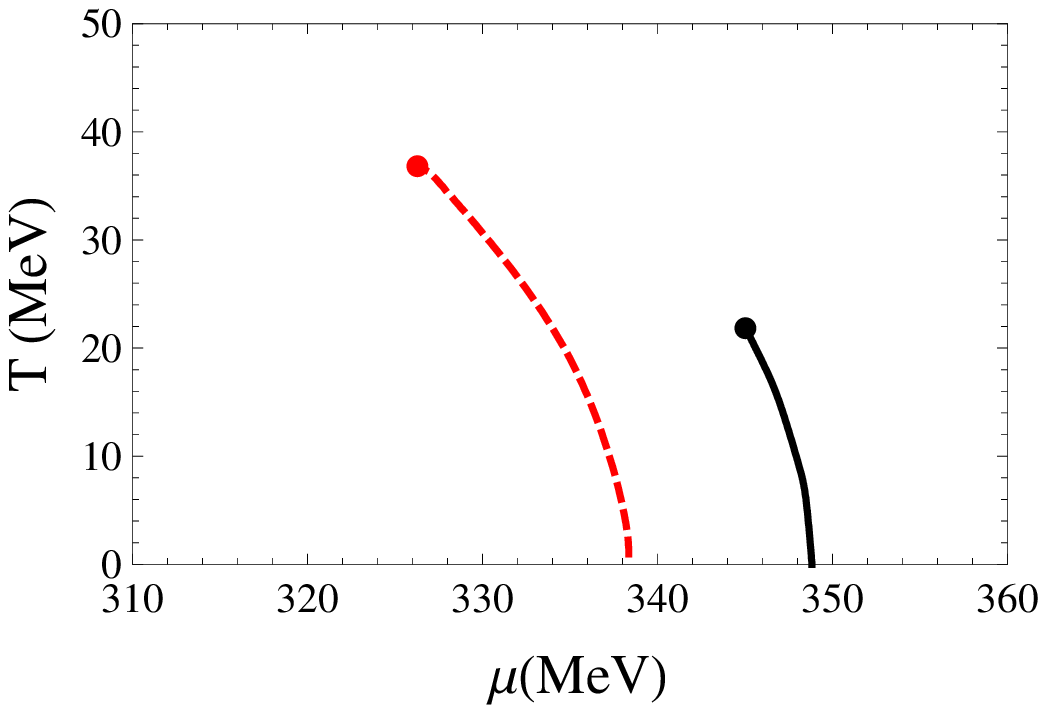,angle=0,width=10cm}
\caption{(color online) Phase diagram in the $T-\mu$ plane.  The continuous
  (OPT) and dashed (LN) lines refer to first order transition lines
  terminating at the CEP which are denoted by the dots.}
\label{phasediag}
\end{figure}

We now consider the effects of both temperature and chemical potential in the
thermodynamics of the NJL model. We start by first  showing the results for
the phase diagram in the $T-\mu$ plane, obtained from both the OPT and the
LN approximations. Away from the chiral limit one expects to have a first order transition
line starting at $T=0$ and at a finite $\mu$, whose value is of the order of
the vacuum quark effective mass, $M_q$. This line vanishes at a critical end
point (CEP) located  at ($\mu_E,T_E$). Then, for $\mu < \mu_E$ and $T > T_E$ a
crossover occurs for finite $m_c$ values. {}Figure \ref{phasediag} displays
the LN and OPT results, illustrating the situation. The OPT critical end point
occurs at $T_E \simeq 22 \rm {\rm MeV}$ and $\mu_E  \simeq  345 \rm {\rm
  MeV}$, while the LN approximation prediction is  $T_E  \simeq 37 \rm {\rm
  MeV}$ and $\mu_E \simeq 326  \rm {\rm MeV}$ (note that previous results for
the critical end points in the NJL model at LN have also been calculated
in e.g. Refs. \cite{barducci}).

{}From {}Fig. \ref{phasediag} we can see that fluctuations brought about by
considering corrections beyond the simple LN approximation can produce quite
large corrections to the critical end points. $T_E$ is about $40\%$ smaller in
the OPT case when compared to the LN, while $\mu_E$ is about $6\%$ higher in
the OPT case. The crossover region is then larger in the OPT case when
compared to the LN results.

We next investigate the usefulness of   the specific heat, the quark
susceptibility and the bulk viscosity, as defined e.g. in \cite{karsch}, as
indicators of the location of the CEP since they all should peak at this
point.  We shall also see how discontinuities arise along the first order
transition line within the interaction measure as defined by
Eq. (\ref{Delta}). The specific heat is given by its standard thermodynamic
definition,

\begin{equation}
C_v= T \frac{\partial s}{\partial T}\;.
\label{Cv}
\end{equation}
The  quark susceptibility can be defined as 

\begin{equation}
\chi_q = \ \frac{\partial \rho}{\partial \mu}\;.
\label{chiq}
\end{equation}
The bulk viscosity is an intrinsic dynamical quantity. However, as shown in
\cite{karsch}, by use of a low-energy theorem of QCD, and assuming  some reasonable
ansatz for the spectral function, it was found that that bulk viscosity can be
expressed in terms of the static thermodynamical quantities derived from the
free energy as~\cite{karsch}

\begin{equation}
\zeta = \frac{1}{9 \omega_0} \left[ T^5 \frac{\partial}{\partial T}
  \frac{(\epsilon-3 P)}{T^4} + 16 | \epsilon_0 | \right]\;,
\label{bulk}
\end{equation}
where $\omega_0$ is a scale where perturbation theory becomes valid and that
here we set it as 
the cutoff $\Lambda$ while $\epsilon_0$ is the  vacuum part of
the energy density. 

\begin{figure}[tbh]
\vspace{0.5cm}  \epsfig{figure=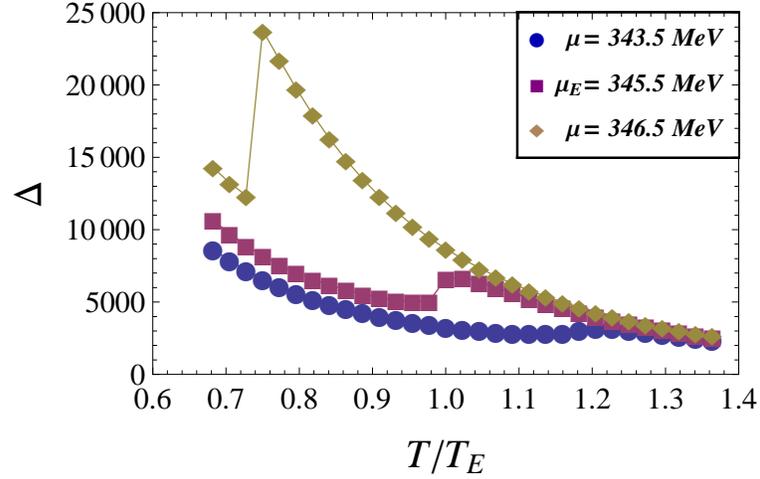,angle=0,width=10cm}
\caption{(color online) OPT results for the interaction measure, $\Delta$, as
  a function of $T/T_{E}$ where $T_{E}$ is the CEP temperature in the OPT
  case. The lines represent situations for $\mu <\mu_{E}$, $\mu=\mu_{E}$ and
  $\mu>\mu_{E}$ where $\mu_{E}$ is the CEP in the OPT case.}
\label{DeltamuT}
\end{figure}

\begin{figure}[tbh]
\vspace{0.5cm}  \epsfig{figure=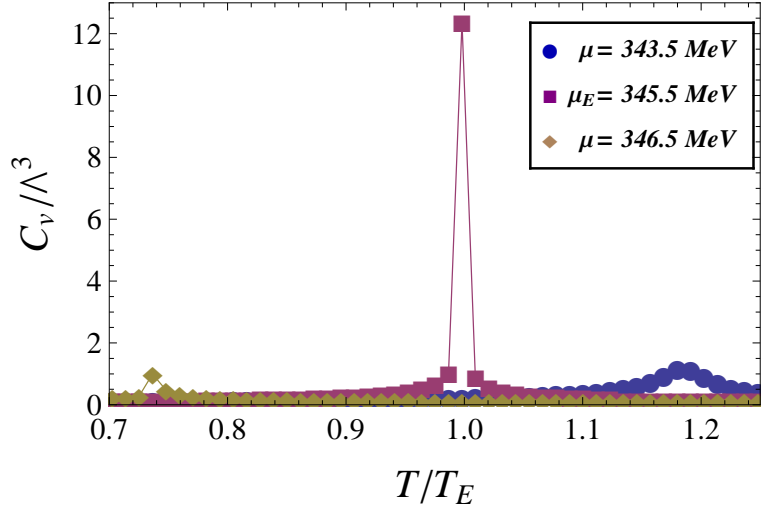,angle=0,width=10cm}
\caption{(color online) OPT results for the dimensionless specific heat,
  $C_v/\Lambda^3$, as a function of $T/T_E$ where $T_E$ is the CEP temperature
  in the OPT case. The lines represent situations for $\mu <\mu_E$,
  $\mu=\mu_E$ and  $\mu>\mu_E$ where $\mu_E$ is the CEP in the OPT case.}
\label{CvmuT}
\end{figure}

\begin{figure}[tbh]
\vspace{0.5cm}  \epsfig{figure=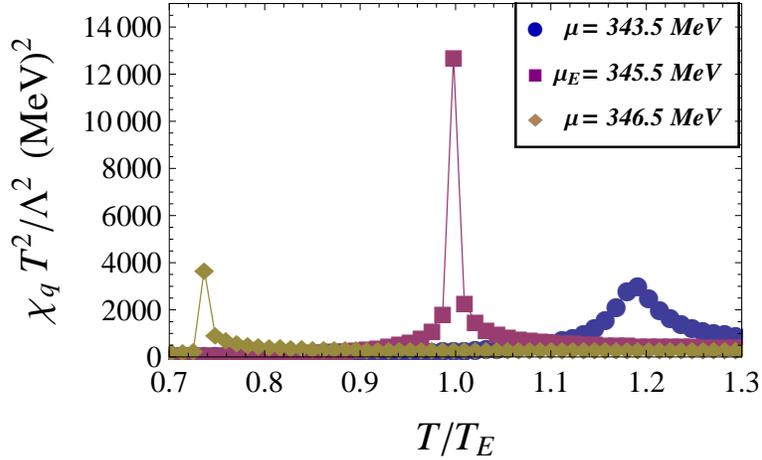,angle=0,width=10cm}
\caption{(color online) OPT results for the normalized quark susceptibility,
  $\chi_q (T/\Lambda)^2$, as a function of $T/T_E$ where $T_E$ is the CEP
  temperature in the OPT case. The lines represent situations for $\mu
  <\mu_E$, $\mu=\mu_E$ and  $\mu>\mu_E$ where $\mu_E$ is the CEP in the OPT
  case.}
\label{chimuT}
\end{figure}

\begin{figure}[tbh]
\vspace{0.5cm}  \epsfig{figure=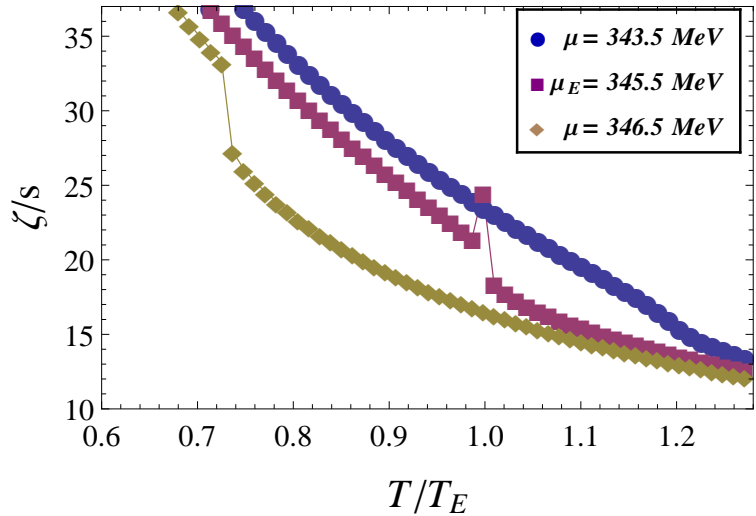,angle=0,width=10cm}
\caption{(color online) OPT results for the bulk viscosity over entropy
  density, $\zeta/s$,  as a function of $T/T_E$ where $T_E$ is the CEP
  temperature in the OPT case. The lines represent situations for $\mu
  <\mu_E$, $\mu=\mu_E$ and  $\mu>\mu_E$ where $\mu_E$ is the CEP in the OPT
  case.}
\label{zetamuT}
\end{figure}

In {}Fig. \ref{DeltamuT} we show the interaction measure as a function of the
temperature at $\mu_{\rm CEP}$ and at values around $\mu_{\rm CEP}$.  The same
is shown in {}Figs. \ref{CvmuT}, \ref{chimuT} and \ref{zetamuT}, for the
specific heat, the quark susceptibility and the ratio of  the bulk viscosity
to the entropy density. In all figures we notice the singular behavior around
the CEP and is much more pronounced for the specific heat and the quark
susceptibility.  We can also see additional structure in all the plots
for values of parameters away from the CEP showing a smooth behavoir for
$T>T_E$ (crossover region) and a discontinuity at smaller temperatures 
(first-order transitions). In its standard definition the interaction measure is
normalized by $T^4$ and therefore the discontinuities are greatly amplified
for the $T<T_E$ situation. This is, for example, the case shown in {}Figs.
\ref{DeltamuT} and \ref{zetamuT} for $\mu=346.5$ MeV (but still smaller  that
$\mu_E$), where the discontinuity happens at the value of the temperature
crossing the critical line shown in {}Fig. \ref{phasediag}.

\section{Conclusions}

We have used the OPT nonperturbative approximation to evaluate Landau's free
energy density for the $SU(2)$ version of the NJL model. By adopting an
adequate form for the interpolation mass parameter we have shown that, in the
large-$N_{c}$ limit, the LN (or MFA)  result is exactly reproduced. At the
first non-trival OPT order, which incorporates a large part of $1/N_c$
corrections but does not involve new parameters beyond those of the LN model,
we have established the consistency with the Goldstone theorem. We have derived
a consistent set of input parameters by matching the model with OPT
corrections to the pion mass and decay constant experimental values, and
obtained consequently OPT deviations from leading-order results in some
quantities, such as typically the GMOR relation.    We then analyzed the cases of
zero temperature and chemical potential, finite temperature and zero chemical
potential, zero temperature and finite chemical potential and for both nonzero
temperature and chemical potential. In each of these cases we have compared
the results for the LN approximation with those from the OPT. 

{}For the finite-temperature, but zero-chemical  potential cases, we have seen
that the inclusion of higher-order corrections to the LN case changes
the LN results only slightly, with changes being at most at about $7\%$. No change of
qualitative  behavior is observed and our results then support the robustness
of  the LN approximation in the studies of the 3+1 dimensional NJL when only
thermal effects are concerned.  However, the inclusion of  density effects
produce results that can lead to a much higher difference between the LN and
when fluctuations are taken into account. This is the case, for example, seen
in determination of the CEP in the phase  diagram of the NJL model at finite
temperature and density. The situation could be antecipated if we recall that,
at first order, the $1/N_c$ corrections considered by the OPT are basically
given  by the square of the scalar density ($\langle {\bar \psi} \psi
\rangle^2$) as well as  the square of the quark number density ($\langle
       {\psi}^+ \psi \rangle^2$). Because of the Dirac  algebra matrices,
       contributions from the scalar and pseudoscalar channels partially
       cancel in the former while they add up for the latter. This is
       interesting since then the OPT can be considered a good alternative to
       the LN approximation in regimes where lattice evaluations become more
       delicate. 

The study of the application of the OPT method to the thermodynamics of
the NJL model made here can also be readily extented to the Polyakov-loop
case of the NJL model. In this case we expect to improve the many
thermodynamical studies made in this context (see e.g. 
\cite{wambach,kenji,chineses} for recent references). We will be reporting
on these results elsewhere.

{}Finally, we have analyzed quantities like the interaction measure (or trace
anomaly), the specific heat, the quark susceptibility and the bulk viscosity
as possible indicators for locating the CEP of QCD, showing  that these
important points  can be located through singular behavior of these functions,
which is more pronounced for the cases of the specific heat and the quark
susceptibility.

\section*{Acknowledgments}
 We thank M. Buballa and S. Descotes-Genon for useful discussions. This work is partially supported
by Conselho Nacional de Desenvolvimento Cient\'{\i}fico e Tecnol\'ogico (CNPq)
and Coordenadoria de Aperfei\c{c}oamente de Pessoal de Ensino  Superior
(CAPES).  M.B.P.  thanks the Nuclear Theory Group at LBNL, UFSC and CAPES for
the sabbatical leave. R.O.R. is partially supported by CNPq and by SUPA, during
the realization of this work in the
United Kingdom.

\appendix

\section{Summing Matsubara frequencies and related formulas}

In this appendix we give the results for the main integrals and Matsubara sums
appearing along the text. The Matsubara sums which are relevant for the
different integrals considered in our work can be derived as (see
e.g. \cite{kapusta})

\begin{equation}
T\sum_{n=-\infty }^{+\infty }\ln [(\omega _{n}-i\mu
  )^{2}+E_{p}^{2}]=E_{p}+T\ln \left[ 1+e^{-\left( E_{p}+\mu \right) /T}\right]
+T\ln \left[ 1+e^{-\left( E_{p}-\mu \right) /T}\right] ,  \label{sum1}
\end{equation}
where $E_{p}^{2}=\mathbf{p}^{2}+\eta ^{2}$. The  zero temperature limit of
Eq.~(\ref{sum1}) is given by 

\begin{equation}
\lim_{T\rightarrow 0}T\sum_{n=-\infty }^{+\infty }\ln [(\omega _{n}-i\mu
  )^{2}+E_{p}^{2}]=E_{p}+\left( \mu -E_{p}\right) \theta (\mu
-E_{p})=\mathrm{max}(E_{p},\mu )\;.  
\label{sum1tzero}
\end{equation}
Likewise, the term,

\begin{equation}
T\sum_{n=-\infty }^{+\infty }\frac{1}{(\omega _{n}-i\mu
  )^{2}+E_{p}^{2}}=\frac{1}{2E_{p}}\left[ 1-\frac{1}{e^{\left( E_{p}+\mu
      \right) /T}+1}-\frac{1}{e^{\left( E_{p}-\mu \right) /T}+1}\right] ,  
\label{sum2}
\end{equation}
gives

\begin{equation}
\lim_{T\rightarrow 0}T\sum_{n=-\infty }^{+\infty }\frac{1}{(\omega _{n}-i\mu
  )^{2}+E_{p}^{2}}=\frac{1}{2E_{p}}\left[ 1-\theta (\mu -E_{p})\right] ;
\label{sum2tzero}
\end{equation}

{}Finally, the finite temperature and density expression

\begin{equation}
T\sum_{n=-\infty }^{+\infty }\frac{\omega _{n}-i\mu }{(\omega _{n}-i\mu
  )^{2}+E_{p}^{2}}=\frac{i}{2}\left [\frac{1}{e^{\left( E_{p}-\mu \right)
      /T}+1}-\frac{1}{e^{\left( E_{p}+\mu \right) /T}+1}\right ]  \;,  
\label{sum3}
\end{equation}
gives, as $T\rightarrow 0$,

\begin{equation}
\underset{T\rightarrow 0}{\lim }T\sum_{n=-\infty }^{+\infty }\frac{\omega
  _{n}-i\mu }{(\omega _{n}-i\mu )^{2}+E_{p}^{2}}=\frac{i}{2}{\rm sgn}\left(
\mu \right) \theta (\mu -E_{p})\;.  
\label{sum3tzero}
\end{equation}

\section{Two-loop calculations for $m_\pi$ and $f_\pi$}

We consider the (first) vertex correction graph to the pion self-energy in
Fig. \ref{fig2loop}  which is not  incorporated by OPT mass insertions within
the one-loop contributions.  Calling its contribution
$\Pi^{(2),ps}_{\pi\:ij}(q^2)$ for the pseudoscalar exchange graphs with
integration momenta $p_1,p_2$ and external momentum $q$, one obtains in
Minkowski space: \be -i \Pi^{(2),ps}_{\pi\:ij}(q^2) = i\frac{\lambda}{2} \int
\frac{d^4p_1}{(2\pi)^4}   \frac{d^4p_1}{(2\pi)^4}  \:Tr[(i\tau_k \gamma_5)
  \frac{i}{p_1{ \hbox{$\!\!\!\!\!/$}}-m}(i\tau_i \gamma_5)  \frac{i}{p_1{
      \hbox{$\!\!\!\!\!/$}}+q{ \hbox{$\!\!\!/$}}-m} (i\tau_k \gamma_5)
  \frac{i}{p_2{ \hbox{$\!\!\!\!\!/$}}+q{ \hbox{$\!\!\!/$}}-m} (i\tau_j
  \gamma_5) \frac{i}{p_2{ \hbox{$\!\!\!\!\!/$}}-m}]\;,
\label{pionv2l}
\ee  while the diagram with scalar $\sigma$ exchange is given by the same
expression with $i\tau_{k,l}\gamma_5 \to 1$ replacements in flavor and Dirac
space.  (Note thus that the $\sigma$-exchange graph has a relative minus sign
with respect to the pion exchange, apart from any other possible factors).
After taking the trace in color, flavor and Dirac spaces, using $\mbox{Tr}\:
\tau_i \tau_j = 2\delta_{ij}$,  we obtain
\begin{eqnarray}
\label{I1pi}
 \Pi^{(2),ps}_\pi(q^2) =&  -4 n_\pi\, N_{\rm f} \lambda \int \frac{d^4p_1}{(2\pi)^4}
 \frac{d^4p_1}{(2\pi)^4} F[p_1] F[p_2] \times \\ \nonumber  &\left[ p_1
   . (p_1+q) p_2.(p_2+q) -m^2 [p_1.(p_1+q) +p_2.(p_2+q)]-m^2 (q^2-m^2)
   \right]\;,
\label{pi2}
\end{eqnarray}
where \be F[p_i] = (p^2_i -m^2)^{-1}[(p_i+q)^2-m^2)]^{-1} \;.\ee Using then
some standard relations like \be \frac{p.(p+q)}{(p^2 -m^2)[(p+q)^2-m^2]} =
\frac{1}{2}\left[\frac{1}{p^2 -m^2}+\frac{1}{(p+q)^2-m^2}\right]
+(m^2-\frac{q^2}{2}) \frac{1}{[(p^2 -m^2)[(p+q)^2-m^2]}  \;,\ee  and upon
  shift of integration momenta\footnote{As usual in such calculations, we
    assume that those manipulations are legitimate under cover of , e.g.,
    Pauli-Villars regularization.}  Eq. (\ref{pi2}) takes, after some algebra,
  the form \be \Pi^{(2),ps}_\pi(q^2) = -8 G N_f N_c n_\pi\, \left[I^2_G(m)
    -q^2 \left(I_G(m) I(q^2)+m^2 I^2(q^2)-\frac{q^2}{4} I^2(q^2)\right)
    \right]\;,
\label{pi2q2ps}
\ee where we defined \be I_G(m) = \int \frac{d^4
  p}{(2\pi)^4}\;\frac{1}{p^2-m^2}\;, \ee which is the standard integral
relevant in the gap-equation, as well as  \be I(q^2) =  \int \frac{d^4
  p}{(2\pi)^4}\;\frac{1}{(p^2-m^2)[(p+q)^2-m^2]}\;,
\label{defI}
\ee which is the standard integral appearing in the one-loop scalar pion
self-energy  contribution Eq. (\ref{pi1}). \\  Calculation steps similar to
those in Eqs. (\ref{pi2}-\ref{pi2q2ps})  give for the scalar $\sigma$ exchange
two-loop  graph contribution in Fig. \ref{fig2loop}): \be \Pi^{(2),s}_\pi(q^2)
= 8 G N_{\rm f} N_c \,  \left[I^2_G(m) -q^2 \left(I_G(m) I(q^2)-m^2
  I^2(q^2)-\frac{q^2}{4} I^2(q^2)\right) \right] \;.\ee
Now combining the scalar and pseudoscalar contributions with the one-loop
contribution (\ref{pi1}), gives for the (inverse) resummed pion propagator the
final expression \bea &  1-2G\,[\Pi^{(1)}(q^2) +\Pi^{(2)}(q^2)] =  1- 4 i G
N_{\rm f} N_c \:\left(2 I_G(m) - q^2 I(q^2)\right) \\ \nonumber  & +16 G^2 N_{\rm f} N_c
\left[(n_\pi-1) \left[I^2_G(m)-q^2 \left(I_G(m)\, I(q^2)-\frac{q^2}{4}
    I^2(q^2)\right)\right] - (n_\pi+1) q^2 m^2 I^2(q^2)\right] \;.
\label{pi2sum}
\eea By inserting the perturbative two-loop gap-equation Eq. (\ref{gap2}) into
the latter expression (in the chiral limit $m_c=0$) one obtains the expected
Goldstone pole at $q^2=0$:  \be 1-2G\,[\Pi^{(1)}(q^2) +\Pi^{(2)}(q^2)] = 4 G
N_{\rm f} N_c\: q^2 \left[ i\, I(q^2) -8G \left[I_G(m)\,
    I(q^2)+\left(2m^2-\frac{q^2}{4}\right) I^2(q^2)\right] \right]\;.  \ee

{}Finally, taking Eq.~(\ref{pi2sum}) with $m_c \ne 0$ and defining the pion
mass as usual as the pole of this expression  for $q^2=m^2_\pi$ gives the
final relation defining $m_\pi$ at this OPT, Eq. (\ref{mpiOPT}). 

By very similar calculations we can derive an expression for the two-loop
vertex contribution to the $\sigma$-meson self-energy, relevant to calculate
the $\sigma$ mass if needed. Contributions are given by graphs similar to
those in Fig. \ref{fig2loop} but with the replacement $i\gamma_5 \tau_{i,j}\to
1$ in flavor and Dirac spaces. After some algebra,  we obtain for the summed
contribution of $\pi, \sigma$ exchanges the expression (at the moment for $m_c
=0$): \be \Pi^{(2),ps+s}_\sigma(q^2) = 8 G N_{\rm f} N_c (n_\pi-1)\,  \left[I^2_G(m)
  +(4m^2-q^2) I_G(m) I(q^2) +\frac{1}{4}\: (4m^2-q^2)^2 I^2(q^2) \right]\;.
\label{sig2q2}
\ee This expression has to be combined with the standard one-loop
contribution: \be \Pi^{(1)}_\sigma(q^2) = 4i G N_{\rm f} N_c \,  \left[I_G(m)
  +\frac{1}{2}\: (4m^2-q^2) I(q^2) \right]\;, \ee to define the inverse
resummed $\sigma$-meson propagator, with $m\to {\cal M}^{OPT}$ for consistency
with the OPT order. Using the gap equation, and solving for the pole of this
propagator gives a (quadratic) equation for $m^2_\sigma$, which explicit
solution is given    in Eq. (\ref{msigma}). 

We next discuss the two-loop contributions to the pion decay constant.
Firstly, at one-loop order, one finds the well-known expression  \be i f^2_\pi
g_{\mu\nu} \delta^{ij} = -\frac{1}{4} \int \frac{d^4 p}{(2\pi)^4} \:Tr[
  \frac{i}{p{ \hbox{$\!\!\!/$}}-m}(i\tau_i \gamma_\mu \gamma_5)  \frac{i}{p{
      \hbox{$\!\!\!/$}}+q{ \hbox{$\!\!\!/$}}-m} (i\tau_j \gamma_\nu \gamma_5)]
\;, \ee where the $1/4$ overall comes from $\tau_i/2$ normalization of axial
currents and the overall minus takes into account trace over fermions. We
adopt  here as a covariant regularization dimensional regularization (at
$D=4$) which is more convenient at two-loop order than Pauli-Villars. Then
using  $\mbox{Tr}\: \tau_i \tau_j = 2\delta_{ij}$, and picking up the
$g_{\mu\nu}$ coefficients  after algebra  gives for $q^2\to 0$ \be f^2_\pi
(\mbox{1-loop}) = -4i N_c m^2 I(0)  \ee which is fully consistent with the
result obtained\cite{klevansky}  from the alternative definition of $f_\pi$
via the one-pion to vacuum transition. 
At two-loop we have a vertex correction graph similar to the first graph in
{}Fig. \ref{fig2loop}  but with the replacement: $ i \gamma_5 \tau_i \to i
\gamma_5 \gamma_\mu \tau_i/2$.  One finds \be i f^2_\pi g_{\mu\nu} \delta^{ij}
=  i\frac{\lambda}{2} \int \frac{d^4p_1}{(2\pi)^4}   \frac{d^4p_1}{(2\pi)^4}
\:Tr[(i\tau_k \gamma_5) \frac{i}{p_1{ \hbox{$\!\!\!\!\!/$}}-m}(i\tau_i/2
  \gamma_\mu \gamma_5) \frac{i}{p_1{ \hbox{$\!\!\!\!\!/$}}+q{
      \hbox{$\!\!\!/$}}-m} (i\tau_k \gamma_5) \frac{i}{p_2{
      \hbox{$\!\!\!\!\!/$}}+q{ \hbox{$\!\!\!/$}}-m} (i\tau_j/2 \gamma_\nu
  \gamma_5) \frac{i}{p_2{ \hbox{$\!\!\!\!\!/$}}-m}]\;.
\label{fpiv2l}
\ee  
By adopting again a Lorentz-covariant preserving regularization (we use dimensional regularization 
but for $D=4$), after some algebra we
obtain the final result \be f^2_\pi (\mbox{2-loop,vertex}) =   8 G N_c
(n_\pi-1) m^4 I^2(0) \;, \ee   where $I(q^2)$ is the one-loop integral as
defined in Eq. (\ref{defI}).  The other remaining two diagrams
in {}Fig. \ref{fig2loop},
of same order, will be again consistently obtained in our case  via the OPT
mass insertion within the one-loop $f_\pi$ expression, so that the final
two-loop expression for $f_\pi$ is the one given in Eq.~(\ref{fpi2OPT}).

\end{document}